\documentclass[11pt,aps,preprint,showkeys,showpacs,amsmath,amssymb,amsfonts,eqnarray,array]{revtex4}
\usepackage{anysize}
\marginsize{1.9cm}{1.9cm}{1.55cm}{1.55cm}
\usepackage{graphicx}
\usepackage{dcolumn}
\usepackage{bm}
\usepackage{setspace}
\begin{document}

%\begin{spacing}{0.9}
%%%%%%%%%%%%%%%%%%%%%%%%%%%%%%%%%%%%%%%%%%%%%%%%%%%%%%%%%%%%%%%%%%%%%%%%%%%%%%%%%%%%%%%%%%%%%%%%%%%%%%%%%%%%%
%\newpage

\title{SURFACE PLASMONS IN COAXIAL METAMATERIAL CABLES\\}
%\vspace{1.0cm}
\author{Manvir S. Kushwaha$^1$ and Bahram Djafari-Rouhani$^2$}
%\vspace{1.0cm} 
\affiliation
{\centerline {$^1$ Department of Physics and Astronomy, Rice University, P.O. Box 1892, Houston, TX 77251, USA}\\
\centerline {$^2$ IEMN, UMR-CNRS 8520, UFR de Physique, University of Science and Technology of Lille I,}
\centerline {59655 Villeneuve d’Ascq CEDEX, France}}

%%%%%%%%%%%%%%%%%%%%%
{\centerline {\em I like to think that no cause is noble if it does not serve mankind as a whole.}}
{\centerline {\hspace{10.7cm} --- Confucius}}
%%%%%%%%%%%%%%%%%%%%%

\date{\today}
%%%%%%%%%%%%%%%%%%%
\begin{spacing}{1.0}
\begin{abstract}
Thanks to Victor Veselago for his hypothesis of negative index of refraction, metamaterials -- engineered
composites -- can be designed to have properties difficult or impossible to find in nature: they
can have both electrical permitivity ($\epsilon$) and magnetic permeability ($\mu$) simultaneously negative.
The metamaterials -- henceforth negative-index materials (NIMs) -- owe their properties to subwavelength
structure rather than to their chemical composition. The tailored electromagnetic response of the NIMs has
had a dramatic impact on the classical optics: they are becoming known to have changed many basic notions
related with the electromagnetism. The present article is focused on gathering and reviewing the fundamental characteristics of plasmon propagation in the coaxial cables fabricated of the right-handed medium (RHM)
[with $\epsilon>0$, $\mu>0$] and the left-handed medium (LHM) [with $\epsilon<0$, $\mu<0$] in alternate
shells starting from the innermost cable. Such structures as conceived here may pave the way to some
interesting effects in relation to, e.g., the optical science exploiting the cylindrical symmetry of the
coaxial waveguides that make it possible to perform all major functions of an optical fiber communication
system in which the light is born, manipulated, and transmitted without ever leaving the fiber environment,
with precise control over the polarization rotation and pulse broadening. The review also covers briefly the nomenclature, classification, potential applications, and the limitations (related, e.g., to the inherent
losses) of the NIMs and their impact on the classical electrodynamics, in general, and in designing the
cloaking devices, in particular. Recent surge in efforts on invisibility and the cloaking devices seems to
have spoiled the researchers worldwide: proposals include not only a way to hide an object without having to
wrap the cloak around it, but also to replace a given object with another, thus adding to the deception even
further! All this is attributed as much to the fundamental as to the practical advances made in the
fabrication and characterization of NIMs. The report concludes addressing briefly the anticipated
implications of plasmon observation in the multicoaxial cables and suggesting future dimensions worth adding
to the problem. The background provided is believed to make less formidable the task of future writers of
reviews in this field.
\end{abstract}
\end{spacing}
%%%%%%%%%%%%%%%%%%%%%%
%\date{\today}
%%%%%%%%%%%%%%%%%%%%%%
\maketitle

\newpage
%%%%%%%%%%%%%%%%%%%%%%%%%%%%%%%%%%%%%%%%%%%%%%%%%%%%%%%%%%%%%%%%%%%%%%%%%%%%%%%%%%%%%%%%%%%%%%%%%%%%%%%%%%%%
\section{A KIND OF INTRODUCTION}

The civilization of the ancient Greeks, Hindus, and Romans has been immensely influential on the art,
education, architecture, language, politics, philosophy, science, and technology of the modern world:
from the surviving fragments of classical antiquity until the Renaissance in Western Europe. While
descriptions of disciplined empirical scientific methods have been employed since the Middle Ages,
the dawn of modern science is generally traced back to the early modern period during what is known
as the Scientific Revolution of the 16th and 17th centuries. From the Middle Ages to the
Enlightenment, the preferred term for the study of nature among English speakers was natural
philosophy. The word scientist, meant to refer to a systematically working natural philosopher (as
opposed to an intuitive or empirically minded one), was coined in 1833 by William Whewell.

As such, people have been curious about the sciences for millennia, albeit the term “physics” was only
coined in the 19th century. In 1850 Cardinal Newman defined it to be: “that family of sciences which
is concerned with the sensible world, with the phenomena which we see, hear, and touch; it is the
philosophy of matter”. Gradually, physicists have narrowed their focus over the past century, perhaps
because they have realized that the span of knowledge has been growing so great and so fast that few,
if any, can encompass it all. Nevertheless, if there is any subject that can still claim to lie at the
heart of knowledge of the natural world, it is physics.

The world's renowned savants -- from the past and present -- believe that the transformation of the
society has produced, to be sure, many beautiful ruins, but not a better society. The reason [for
such disappointment(s)] obviously being the man-made disaster in Hiroshima and Nagasaki during the
World War II! The physics and particularly the solid state physics that has an exceptionally poignant
creation myth is the key-stone behind such a contention.

The solid state physics -- that has changed its name into condensed matter physics to demonstrate its
momentum -- can safely be regarded as the discipline of -ons. These -ons include the “real particles”
such as electron, “quanta of collective excitations” such as plasmon, and “dressed particles” such as
polaron. In the language of the quantum mechanics, these -ons are characterized either as bosons or as
fermions -- the former obey the Bose-Einstein statistics whereas the latter the Fermi-Dirac statistics.
The name and the underlying concept associated with the -ons (anion, boson, electron, exciton, fermion,
helicon, hoctron, magnon, neutron, phonon, photon, plasmon, plasmeron, polariton, polaron, proton,
roton, skyrmion, ...) were, often, the work of separate people, far removed in time. These -ons are
to the wave-propagation characteristics as are the steel girders and concrete to the modern structures.

Our interest here is in the (surface) plasmon: the collective excitation of electron density bound to
the surface. The free electron gas in metals and electron-hole gas in semiconductors make up the solid
state plasmas. Depending upon the wavelength of the probe, the solid state plasma can reveal two very
different modes of behavior: single-particle excitations and collective (plasmon) excitations. The two
domains of behavior are distinguished by the critical length ($\lambda_c$) called Debye (screening)
length in classical (quantum) plasmas. For $\lambda > \lambda_c$ ($\lambda < \lambda_c$) the plasma
responds collectively (single-particle-like). The surface plasmon is a well-defined excitation that
can exist on an interface that separates a surface-wave active medium (with $\epsilon < 0$) from a
surface-wave inactive (with $\epsilon > 0$) medium. It is characterized by the electromagnetic fields
that are localized at and decay exponentially away from the interfaces. In a conventional system, and
under the normal physical conditions, an interface supports one and only one confined mode associated
with either p-polarization or s-polarization [1-16].

%fig. 1
\begin{figure}[htbp]
\includegraphics*[width=7cm,height=7cm]{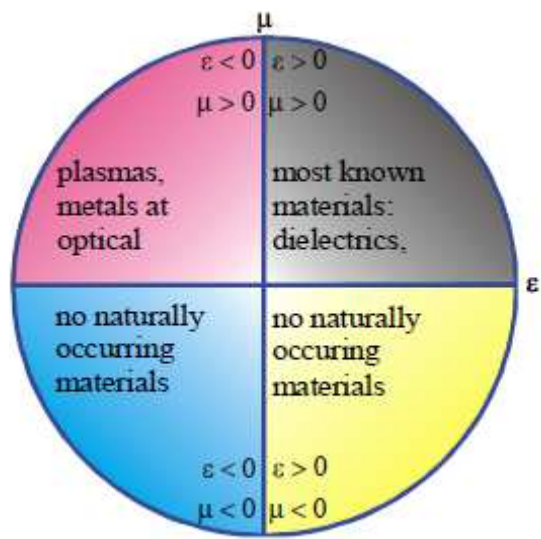}
%%%%%%%%%%%
\caption{(Color online) This picture shows the four possible combinations of $\pm \epsilon$, $\pm \mu$ .
In each quadrant new phenomena are observed. For example when $\epsilon < 0$ surface plasmons are
observed. Therefore it is as if a new door has been opened into ‘third quadrant’ electromagnetism. One
of the phenomena seen here is negative refraction, but there are other remarkable effects to be found
such as subwavelength imaging. (After J.B. Pendry, Ref. 121).}
\label{fig1}
\end{figure}

It is now widely known that mid 1970s had begun to offer the condensed matter physicists with quite new
and exciting venues to explore: the semiconducting quantum structures with reduced dimensions such as
quantum wells, quantum wires, quantum dots, and their periodic counterparts. The tremendous research
interest in the quantum phenomena associated with these systems was obviously spurred by the discovery
of quantum Hall effects -- both integral [17] and fractional [18]. While this momentum still seems to be
growing [19], the classical phenomena emerged with the proposal of the photonic (and phononic) crystals
[20-22]  and, more recently, the negative-index metamaterials (NIMs) have been drawing considerable
attention of numerous research groups worldwide. Conceived and hypothesized some four decades ago by
Veselago [23], theorized through the proposal of superlens by Pendry [24], and first practically realized
by Smith and co-workers [25], an artificially designed negative-index metamaterial -- exhibiting
simultaneously negative electrical permittivity $\epsilon (\omega)$ and magnetic permeability
$\mu (\omega)$ and hence negative refractive index $n = \pm \sqrt{\epsilon \mu}$ [see Fig. 1] -- seems to
have changed many basic notions of the traditional (or conventional) electromagnetism.

%fig. 2
\begin{figure}[htbp]
\includegraphics*[width=8cm,height=9cm]{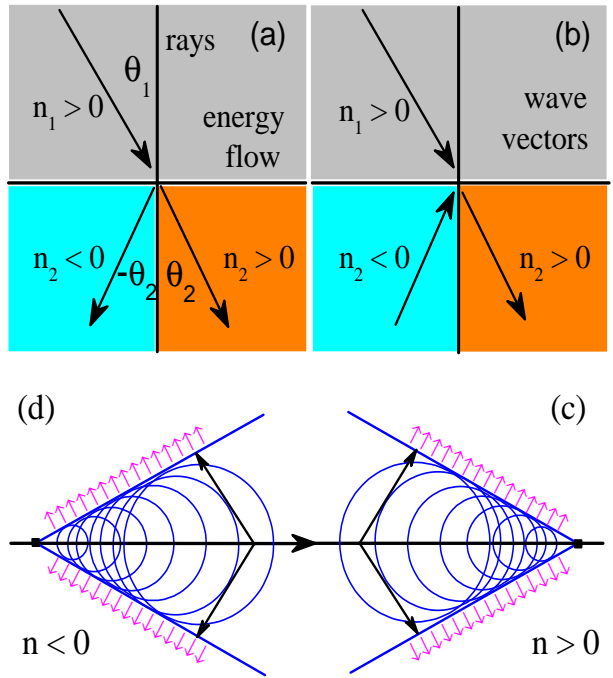}
%%%%%%%%%%%
\caption{(Color online) A material with negative refractive index ($n < 0$) bends rays of incident light to
the ‘wrong’ side of the normal (a). Meanwhile the wave vector points in the opposite direction to the energy
flow because the group velocity is negative (b). In the conventional system with ($n > 0$) the Cherenkov
radiation is emitted from the cone in front of the charged particle (c), whereas in the NIMs ($n < 0$) it is
emitted from the cone behind the charged particle (d).}
\label{fig2}
\end{figure}

A superlens (or perfect lens) is a lens which is made up of NIMs to go beyond the {\em diffraction limit}.
The diffraction limit is an inherent barrier due, in part, to the imperfections in the lenses or
misalignment in the conventional optical devices. There is a fundamental maximum to the resolution of any
optical system which is due to the diffraction limit. An optical system with the ability to produce images
with angular resolution as good as the device's theoretical limit is said to suffer from the diffraction
limit. The observation of sub-wavelength structures with microscopes is a {\em hard nut to crack} because
of the (Abbe) diffraction limit. In 1873, Ernst Abbe found that light with wavelength $\lambda$ traveling
in a medium with refractive index $n$ and converging to a spot with angle $\theta$ will make a spot with
diameter $d=\lambda/[2 \, n \,{\rm sin} (\theta)]$. The factor $n \,{\rm sin} (\theta)$ in the denominator
is called the {\em numerical aperture} and can reach about 1.25 in modern optics, hence the Abbe limit turns
out to be $d \approx \lambda/2$. The  green light with $\lambda \approx 500$ nm yields the Abbe limit close
to 200 nm, which is large compared to most nanostructures or biological cells which have sizes on the order
of 1 $\mu$m. In 2000, John Pendry proposed a superlens to be fabricated of metamaterial that was shown to
compensate for the wave decay and to reconstruct images in the near field [24]. While theory and simulations
demonstrate that the superlens and hyperlens can, in principle, work, engineering obstacles need to be
overcome.

%fig. 3
\begin{figure}[htbp]
\includegraphics*[width=8cm,height=7cm]{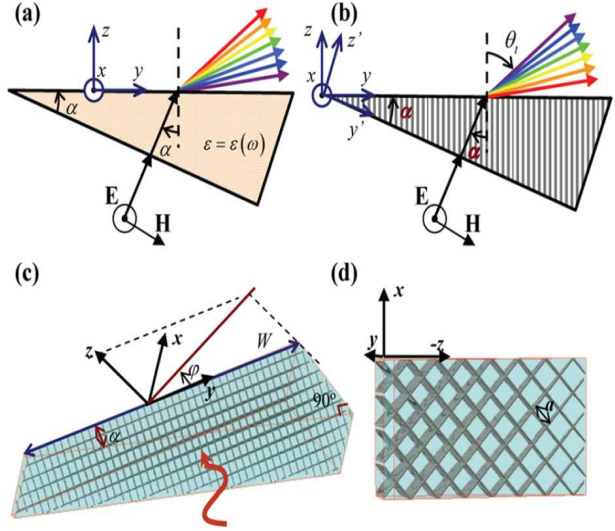}
%%%%%%%%%%%
\caption{(Color online) (a) Refraction of light by a conventional dielectric prism. (b) Refraction of light
by the metamaterial prism proposed in this Letter. Notice that the palette of refracted colors is reversed.
(c) and (d) Perspective views of the metamaterial prism. The metamaterial is formed by nonconnected crossed
metallic wires that lie in planes normal to the y direction. (After M.G. Silveirinha, Ref. 58).}
\label{fig3}
\end{figure}

A NIM forms a left-handed medium (LHM), with the energy flow ${\boldsymbol E} \times {\boldsymbol H}$ being
opposite to the direction of propagation, for which it has been argued that such phenomena as Snell’s
law, Doppler effect, and Cherenkov radiation are inverted. What actually does that mean? Let us take, for
example, Snell’s law [$n_1$ sin ($\theta_1$)$=n_2$ sin ($\theta_2$)]:  (i) in the conventional system,
light is bent towards the normal with $n_2 > n_1$ and $\theta_2 < \theta_1$ [see Fig. 2 (a)], but (ii) in
the left-handed metamaterials, (with $n_2 < 0$), light is bent on the same side of the normal with
$sin (\theta_2) < 0$ [see Fig. 2 (a)]. The time-averaged Poynting vector is antiparallel to phase velocity.
However, for waves (energy) to propagate, a -$\mu$ must be paired with a -$\epsilon$ in order to satisfy
the wave number dependence on the material parameters $k = \omega\sqrt{\epsilon \mu}$ [see Fig. 2 (b)].
Cherenkov radiation is emitted when a charged particle [such as an electron] travels through a dielectric
(electrically polarizable) medium with a speed greater than that at which light would otherwise propagate
in the same medium. As the charged particle travels, it disrupts the local electromagnetic field in the
medium thereby polarizing the atoms in the dielectric. Photons are emitted as an insulator's electrons
restore themselves to the ground state after the disruption has passed. A reverse Cherenkov effect can be
experienced using the NIMs: this means that when a charged particle passes through a (metamaterial) medium
at a speed greater than the speed of light in that medium, that particle will radiate from a cone behind
itself [see Fig. 2 (d)], rather than in front of it (as is the case in the conventional materials)
[see Fig. 2 (c)].

%fig. 4
\begin{figure}[htbp]
\includegraphics*[width=7.5cm,height=8cm]{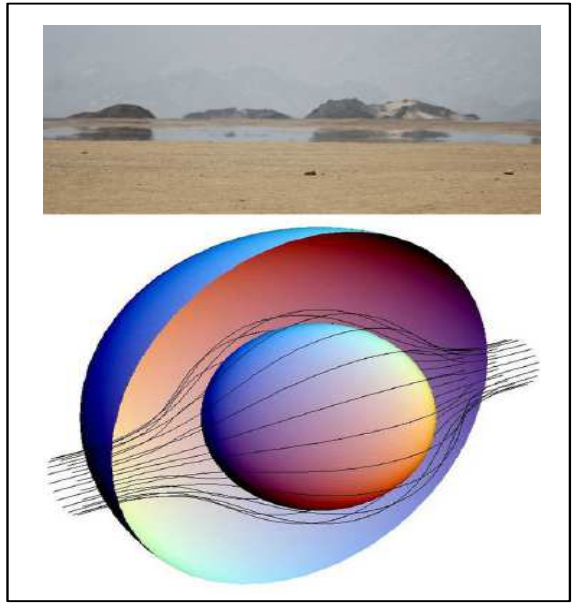}
%%%%%%%%%%%
\caption{(Color online) (Top panel) A hot desert surface causes a refractive index gradient in the air
above, causing rays of light to be refracted continuously to form a reflection in the road and hence
the appearance of water: mirage. (Bottom panel) Similarly, a graded refractive index cloak can guide
light around a hidden object so that an observer sees only that which is behind the cloak. (After J.B.
Pendry, Ref. 67).}
\label{fig4}
\end{figure}

The subject of metamaterials, or (artificially) engineered composites, has gained un unexpected momentum
and the research interest seems to have focused not only on the photonic crystals with metamaterial
components [26-33] but also on the single- and multi-layered planar structures [34-43] as well as on the
(usually) single cylindrical geometries [44-55]. The interesting phenomena emerging from the geometries
involving metamaterials include the slowing, trapping, and releasing of the light signals [56], the
proposal of the cloaking devices [57], and the extraordinary refraction of light [58] (see Fig. 3). The
early development of the subject can be found in interesting review articles by Pendry [59], by Boardman
[60], and by Shalaev [61]. Cloaking is an illusion like a mirage: you steer light around an object and
therefore you never see the object [see Fig. 4]. The practical interest in devising the cloaking devices
is no longer confined to the electromagnetic waves [62-72], but has also imbued the similar quest related
with the acoustic waves [73-85] and the plasmonic waves [86-98]. The {\em cloakmania} has gone to the
extent that some scientists think that all the attention paid to cloaking phenomenon distracts us from the
technology's true potential: anything that makes use of the electromagnetic spectrum might be improved or
altered! The tailored response of the metamaterials has had a dramatic impact on engineering, material
science, optics, and physics communities alike, because they can offer electromagnetic properties that are
difficult or impossible to achieve with naturally occurring (conventional) materials.

%fig. 5
\begin{figure}[htbp]
\includegraphics*[width=7.5cm,height=8cm]{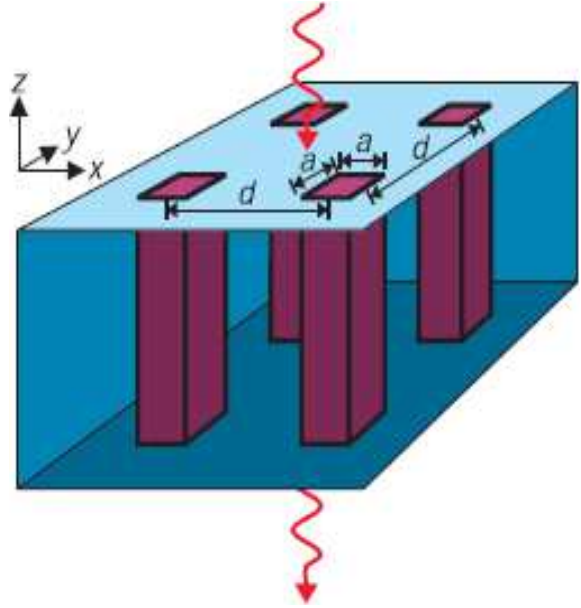}
%%%%%%%%%%%
\caption{(Color online) The model system: $a \times a$ square holes arranged on $d \times d$ lattice are cut
into the surface of a perfect conductor. The theory predicts localized surface plasmon modes induced by the
resultant structure. (After J.B. Pendry et al., Ref. 103).}
\label{fig5}
\end{figure}

The recent research interest in surface plasmon optics has been invigorated by the experiment performed on
the transmission of light through sub-wavelength holes in metal films [99] (see Fig. 5). This experiment
has spurred numerous theoretical [100-104] as well experimental [105-110] works on similar structured
surfaces: either perforated with holes, slits, dimples, or decorated with grooves. It has been argued that
resonant excitation of surface plasmons creates huge electric fields at the surface that force the light
through the holes, yielding very high transmission coefficients. The idea of tailoring the topography of a
perfect conductor to support the surface waves resembling the behavior of surface plasmons at optical
frequencies was discussed in the context of a surface with an array of two-dimensional holes [104]. The
experimental verification of this proposal has recently been reported [111-113] on the structured
metamaterial surfaces which support surface plasmons at microwave frequencies. Because of their mimicking
characteristics, these geometry-controlled surface waves were named {\em spoof} surface plasmons [see Fig.6].

%fig. 6
\begin{figure}[htbp]
\includegraphics*[width=7.5cm,height=8cm]{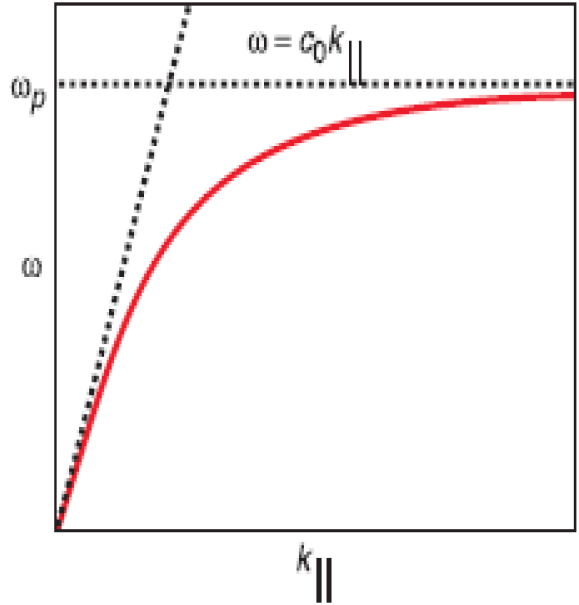}
%%%%%%%%%%%
\caption{(Color online) The dispersion relation for the spoof surface plasmons on a structured surface. The
asymptotes of the light line at low frequencies and the plasma frequency at large values of $k_{\parallel}$
are shown. Note that at large $k_{\parallel}$ the frequency of the mode approaches $\omega_p$, in contrast
to an isotropic plasma where the asymptote is $\omega_p /\sqrt{2}$. (After J.B. Pendry et al., Ref. 103).}
\label{fig6}
\end{figure}

Here, it would be interesting to shed some light on how the plasma frequency is lowered in the metamaterials
structured periodically with wire loops or coils [see, e.g., Fig. 7]. Some time ago, Pendry and coworkers
[114] argued that any restoring force acting on the electrons will not only have to work against the rest
mass of the electrons, but also against the self-inductance of such wire structures. This effect is of
paramount importance in these wire structures. They went on arguing that the inductance of a thin wire
diverges logarithmically with wire radius and confining the electrons to thin wires enhances their
effective mass by orders of magnitude. In other simpler format [105] one can, from Ohm's law
($j=\sigma E_{local}$), determine the effective conductivity for the inductive wire , and calculate an
effective local dielectric function analogous to the Drude dielectric function, but with plasma frequency
directly related to the inductance ($L$) of the unit cell (of length $l$) and wire spacing $d$ according
to $\omega_p=\sqrt{l/(d^2 L \epsilon_0 )}$. Thus reducing the wire radius enhances the inductance which
thereby lowers the plasma frequency of the system. Such estimates led them to predict the plasma frequency
on the order of $\sim$ 7 to 8 GHz.

%fig. 7
\begin{figure}[htbp]
\includegraphics*[width=7cm,height=9cm]{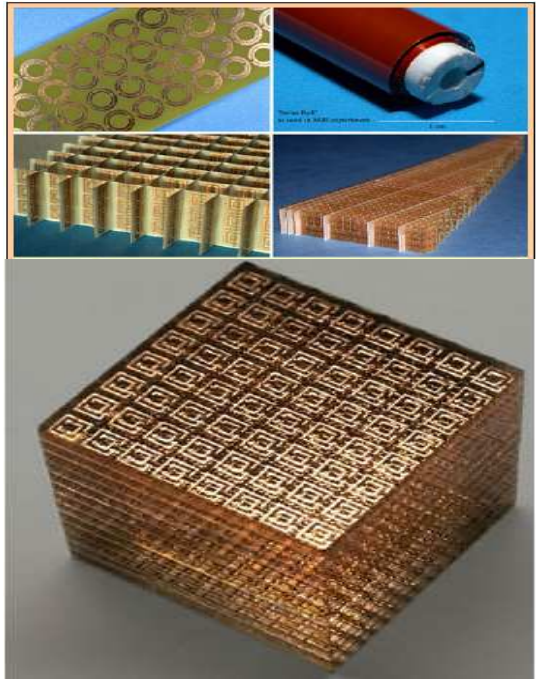}
%%%%%%%%%%%
\caption{(Color online) Top panels: Examples of metamaterials used in the microwave experiments. The unit
cells are of the order of 5mm across. Bottom panel: ‘The Boeing cube’-- a structure designed for the
negative refractive index in the GHz range. (After J.B. Pendry and D.R. Smith, Ref. 59).}
\label{fig7}
\end{figure}

The purpose of this article is to review the recent research efforts dedicated to investigate the surface
plasmon propagation in the coaxial cables fabricated of metamaterials interlaced with conventional
dielectrics using the Green-function (or response function) theory in the absence of an applied magnetic
field [115-116]. Theoretical framework of the Green function theory has already been successfully tried
and tested for the conventional (semiconducting) coaxial cables both with and without an applied magnetic
field [117-119]. To be explicit, we discuss the propagation characteristics of surface plasmons in coaxial
cables made up of right-handed medium (RHM) [with $\epsilon >0$, $\mu >0$] and the left-handed medium (LHM)
[with $\epsilon(\omega) <0$, $\mu(\omega) <0$] in alternate shells starting from the innermost cable. In
other words, we visualize a cylindrical analogue of a one-dimensional planar superlattice structure bent
round until two ends of each layer coincide to form a multicoaxial cylindrical geometry. We prefer to name
such a resultant structure as multicoaxial (metamaterial) cables.

Such structures as conceived here may pave the way to some interesting effects in relation to, for example, the
optical science exploiting the cylindrical symmetry of the coaxial waveguides that make it possible to perform
all major functions of an optical fiber communication system in which the light is born, manipulated, and
transmitted without ever leaving the fiber environment, with precise control over the polarization rotation and
pulse broadening [120]. The cylindrical geometries are already known to have generated particular interest for
their usefulness not just as electromagnetic waveguides, but also as atom guides, where the guiding mechanism
is governed mainly by the excited cavity modes. It is envisioned that the understanding of atom guides at such
a small scale would lead to much desirable advances in atom lithography, which in turn should facilitate atomic
physics research [121].

There is one trait that many theoretical physicists share with philosophers. In both cases the interest in a
field of study seems to vary in inverse proportion to how much one must learn to qualify as an expert. There
is an important core of truth to the fact that ``expertness is what survives when what has been learnt has
been forgotten". True learning leading someone to become an expert is a long-term project -- ideally,
life-long. However, present day learning of any subfield of science, particularly the vastly growing physics,
motivated to gain mastery is inconceivably precarious for survival in the existing competitive world of
science. In this relatively short space we shall try to review briefly what we have been able to learn from
theoretically motivated analytical diagnoses of the surface plasmon propagation in the coaxial metamaterial
cables in the cylindrical symmetry. We appeal to the interested readers to digest what they can and blame the
authors for the rest.

The rest of the paper is organized as follows. In Sec. II, we discuss some basic notions of the theoretical
framework employed to derive exact inverse response functions within the framework of Green function theory.
In Sec. III, we report several interesting illustrative examples on the plasmon dispersion and density of
states in a variety of experimentally feasible situations for single-, double-, and multiple-interface coaxial
cables. In Sec. IV, we list some potential technological applications of the metamaterials in diverse shapes,
sizes, and dimensions. Finally, we conclude our findings and suggest some interesting dimensions worth adding
to the problem in Sec. V.

%%%%%%%%%%%%%%%%%%%%%%%%%%%%%%%%%%%%%%%%%%%%%%%%%%%%%%%%%%%%%%%%%%%%%%%%%%%%%%%%%%%%%%%%%%%%%%%%%%%%%%%%%%%%%%%%%%%
\section{STRATEGY OF THE GREEN-FUNCTION THEORY}

Recently, we have embarked on a systematic investigation of the surface plasmon excitations in the coaxial
cylindrical shells made up of left-handed metamaterials interlaced with right-handed media within the
framework of the Green-function (or response function) theory (GFT) [115-116]. The knowledge of such
excitations is fundamental to the understanding of the plasmon optics in the system. We consider the
cross-section of these coaxial cables to be much larger than the de-Broglie wavelength, so as to neglect
the quantum size effects. We include the retardation effects but neglect, in general, the damping effects
and hence ignore the absorption. In the state-of-the-art high quality systems, this is deemed to be quite
a reasonable approximation [25]. Thus we study the plasmon excitations in a neat and clean system
comprised of coaxial metamaterial cables [see Fig. 8]. While it is always important to have a paper as much
self-contained as possible, we think that reiterating all the mathematical details from Ref. [115] would
make it an unwanted repetition. That is why we choose to give here only a sketch of the strategy of working
within the GFT and refer the reader for the details to Ref. 115.

%fig. 8
\begin{figure}[htbp]
\includegraphics*[width=7.5cm,height=8cm]{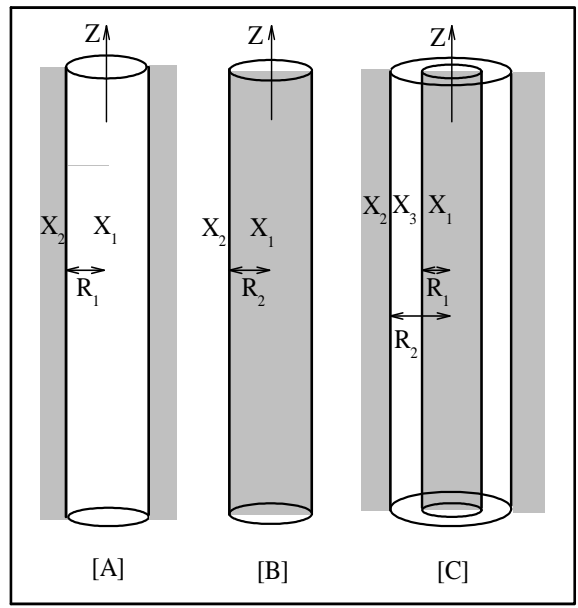}
%%%%%%%%%%%
\caption{(Color online) Schematics of the concept of three perturbations: [A], [B], and [C]. The blank
(shaded) region refers to the material medium (black box) in the system. The sum of the first two
perturbations defines a metamaterial (dielectric) cylinder embedded in a dielectric (metamaterial) and
the sum of all three perturbations specifies a metamaterial (dielectric) shell surrounded by two
unidentical dielectrics (metamaterials). Here $R_j$ is the radius and $X\equiv \epsilon (\omega)$ or
$\mu (\omega)$ for a specific medium. (After Kushwaha and Djafari-Rouhani, Ref. 115).}
\label{fig8}
\end{figure}

We consider the plasma waves propagating with an angular frequency $\omega$ and the wave vector $\vec{k}$
[$\parallel \hat{z}$ in a medium with cylindrical symmetry ($\rho,\theta, z$). The plasma waves will be
throughout assumed to observe the spatial localization in the plane perpendicular to the axis of the
cylinder. Note that the situation here is totally different from the Cartesian co-ordinate system in which
one can readily define a sagittal plane (i.e., the plane defined by the wave vector and the normal to the
surface/interface) and hence isolate the transverse magnetic (TM) and the transverse electric (TE) modes,
at least in the absence of an applied magnetic field. The only exception to this notion is the Voigt
geometry (with a magnetic field parallel to the surface/interface and perpendicular to the propagation
vector) that can still (i.e., even in the presence of an applied magnetic field) allow the separation of
the TM and TE modes (see, for details, Ref. 19). In the literature, the TM and the TE modes are also
referred to as p-polarization and s-polarization, respectively.

It is noteworthy that the roots of our Green-function theory lie virtually in the interface-response
theory (IRT) [123] generalized to be applicable to such quasi-one dimensional (Q1D) systems as
investigated here [117]. Ever since its inception, the IRT has been extensively applied to study various
quasi-particle excitations such as phonons, plasmons, magnons, ...etc. in the heterostructures and
superlattices [see, e.g. Ref. 19]. Lately, it has also been successfully applied to study the magnonic,
phononic and photonic band-gap crystals.

Before we proceed further, it is crucial to define a characteristic feature of the IRT: the black-box
surface (BBS). By BBS we mean an entirely opaque surface through which electromagnetic fields cannot
propagate. The idea of introducing the BBS in the IRT was conceived with two prominent advantages over
the contemporary semiclassical approaches in mind. First, it allows one to disconnect completely from
the extra mathematical world and hence to confine only within the truly building block of the system
concerned. Second, it provides a great opportunity to get rid of using the boundary conditions one is
so routinely accustomed to in dealing with the inhomogeneous systems. What results is a number of
simplified and compact forms of the response functions which one only needs to sum up in order
to proceed further for studying the desired physical property of the resultant system. Conceptually,
this is achieved by imposing that $c$ (the speed of light in vacuum), $\epsilon$ (the electric
permittivity), and  $\mu$ (the magnetic permeability) vanish inside a specific region. In order to
create a medium bounded by a black-box surface, we assume that Eqs. $(2.5)-(2.8)$ [in Ref. 115] are
only valid for either $\rho > R$ or $\rho < R$, with $R$ as the radius of the only cylinder in question
by now. Then we multiply the right-hand sides of Eqs. $(2.5)-(2.8)$ [in Ref. 115] by the step function
$\theta(\rho-R)$ or $\theta(R-\rho)$, as the case may be.

It should be noted first, that all the quantities referred to earlier or to be referred in what follows
will carry a subscript $j$ when referring to a given perturbative operation. The first and the foremost
point is to create a black-box surface in order to confine within the building block of the system and
disconnect altogether from the rest of the mathematical world. For this purpose, we assume a step
function $\theta (...)$ specifying a given physical situation in front of field components such as given
in Eqs. $(2.5)-(2.8)$ in Ref. 115, for example. This then leads us to define a cleavage operator
$\tilde{V}_j(...)$, which is, in fact, a $2n \times 2n$ matrix, where $n$ is the number of interfaces in
the problem. Now we also know beforehand that there is a bulk Green's function matrix $\tilde{G}_j (...)$
representing the medium we are confined to. With this, we define a response operator
%eq. 1
\begin{equation}
\tilde{A}_j(...) = \tilde{V}_j(...)\,\tilde{G}_j(...)
\end{equation}
The arguments of all of these matrices are many depending upon the physical problem at hand, but the two
which are the most important to be specified are $\rho$ and $\rho'$ in the present problem. Evidently, the
response operator is also a $2n \times 2n$ matrix. Next, we define an operator
%eq. 2
\begin{equation}
\tilde{\Delta}_j(...) = \tilde{I} + \tilde{A}_j(...)
\end{equation}
where $\tilde{I}$ is a unit matrix of the same order as the rest. Now we need to calculate the inverse of
the bulk Green's function $\tilde{G}_j(...)$, which is given by, say, $\tilde{G}^{-1}_j(...)$. As such, we
now have all what we need to calculate the inverse response function $\tilde{g}^{-1}_j(...)$ in the
interface space (say, $M_s$). This is defined by
%eq. 3
\begin{equation}
\tilde{g}^{-1}_j(...) = \tilde{\Delta}_j(...)\,\tilde{G}^{-1}_j(...)
\end{equation}
Notice that $\tilde{g}^{-1}_j(...)$ represents {\it exclusively} the response function of the region we are
initially confined to rather than for the physical system we may have been interested in. To be more
explicit, suppose that $\tilde{g}^{-1}_1(...)$ in Eq. (3) represents the dielectric, metallic, or
semiconducting cylinder surrounded by a black-box. And suppose we are interested in a physical system made
up of this cylinder surrounded by some real, but different, material. Then we will have to follow the steps
identical to those leading to Eq. (3) but now confining to the semi-infinite region enclosing the black-box.
Suppose the latter system turns out to be represented by some inverse response function
$\tilde{g}^{-1}_2(...)$. Then the final physical system made up of a semiconducting cylinder surrounded by a
dielectric is represented by
%eq. 4
\begin{equation}
\tilde{g}^{-1}_f(...) = \tilde{g}^{-1}_1(...) + \tilde{g}^{-1}_2(...)
\end{equation}
This response function serves many useful purposes in the realistic situations. For instance, the determinant
of $\tilde{g}^{-1}_f(...)$ equated to zero yields the respective dispersive modes of, e.g., a semiconducting
cylinder surrounded by a semi-infinite dielectric. It also becomes useful to calculate the local as well as
total density of states [115]. Analogous response functions are also useful to compute numerous electronic,
and optical properties of a given system under different physical conditions. Such is the strategy of working
within our GFT employed to obtain the results discussed in what follows.

The use of the GFT has numerous advantages over the traditional Maxwell equations with the boundary conditions.
It is quite well known that the Maxwell equations with proper boundary conditions only provide us with the
dispersion relations for the electromagnetic waves in an inhomogeneous medium. The GFT, which is essentially a
matrix formulation, on the other hand, not only enables us to obtain the dispersion relations for the desired
excitations but also allows us to study various static and dynamic properties in terms of the response function
for the resultant system. These include, e.g., the local and total density of states, reflection and
transmission coefficients, inelastic electron and light scattering, tunneling phenomena, and selective
transmission, to name a few [19].

\section{ILLUSTRATIVE EXAMPLES}

As can be seen in Ref. 115, our final analytical results for the dispersion characteristics are Eqs. (3.26) and
(3.32), respectively, for the single cylindrical cable embedded in some different material background and the
coaxial cylindrical system made up of a finite shell bounded by the closed (innermost) cable and the
semi-infinite medium. Note that both of these equations are, in general, the complex transcendental functions.
Therefore, in principle, we need to search the zeros of such complex functions. We had to strike a compromise
among a few choices. We decided to ask the machine to produce those zeros where the imaginary (real) part of the
function changes the sign, irrespective of whether or not the real (imaginary) part is zero in the radiative (nonradiative) region.  We believe this has resulted into a reliable scheme for studying the dispersion
characteristics of plasmons in the present systems. This is because all the plasmon modes (confined or extended)
are found to have excellent correspondence with the peaks in the local and/or total density of states. We
purposely consider only the cases with dispersive metamaterials interlaced with conventional dielectrics
(usually vacuum with $\epsilon =1=\mu$).
%The detailed studies of nondispersive metamaterials is deferred to a future publication.
So the only parameters involved in the treatment are $F$ and $\omega_0$ and we choose them such that $F=0.56$, $\omega_p/2\pi=10$ GHz and $\omega_0/2\pi=4$ GHz; the latter yields $\omega_0/\omega_p=0.4$
[see Ref. 115]. We will later assign an additional numeral as a suffix to the background dielectric constants
corresponding to the region in the geometry concerned. Other parameters such as the ratio of the radii of the
cylinders $R_2/R_1$, the normalized plasma frequency $\omega_p R/c$, and the azimuthal index of the Bessel
functions $m$ will be given at the appropriate places during the discussion. We will present our results in terms
of the dimensionless propagation vector $\zeta=ck/\omega_{p}$ and frequency $\xi=\omega/\omega_{p}$, where
$\omega_p$ stands for the screened plasma frequency. Both local and total DOS will be shown in arbitrary units throughout.

\subsection{Single-interface systems}

%fig. 9
\begin{figure}[htbp]
\includegraphics*[width=7.5cm,height=8cm]{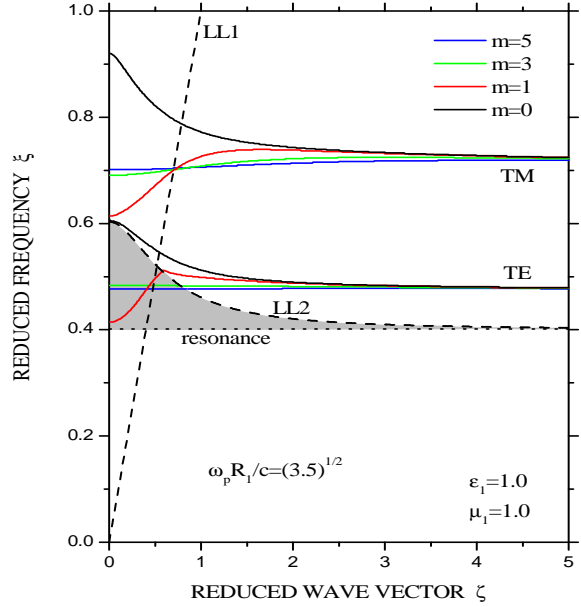}
%%%%%%%%%%%
\caption{(Color online) Plasmon dispersion for a dielectric (vacuum) cable embedded in a metamaterial background
for different values of index $m=0$, 1, 3, 5. The dimensionless plasma frequency used in the computation is
specified by $\omega_p R_1/c=\sqrt{3.5}$. Dashed line and curve marked as LL1 and LL2 refer, respectively, to the
light lines in the vacuum and the metamaterial. The horizontal dotted line stands for the characteristic
resonance frequency ($\omega_0$) in the metamaterial. The shaded area represents the region within which both
$\epsilon(\omega)$ and $\mu(\omega)$ are negative and prohibits  the existence of the confined modes.
(After Kushwaha and Djafari-Rouhani, Ref. 115).}
\label{fig9}
\end{figure}

Figure 9 illustrates the plasmon dispersion for the dielectric (vacuum) cable embedded in a metamaterial background
for $m=0$, 1, 3, and 5. The plots are rendered in terms of the dimensionless frequency $\xi$ and the dimensionless propagation vector $\zeta$.  The important parameter involved is the dimensionless plasma frequency specified by $\omega_pR_1/c=\sqrt{3.5}$. The dashed line and the curve marked as LL1 and LL2 refer, respectively, to the light
lines in the vacuum and the metamaterial. The horizontal dotted line stands for the characteristic resonance
frequency ($\omega_0=0.4 \omega_p$) in the metamaterial. The shaded area represents the region where
$\epsilon(\omega)< 0$ and $\mu(\omega)< 0$ and disallows the existence of the confined modes [see Sec. III.G in
Ref. 115]. It is important to notice that the confined modes can be distinguished as TM or TE only when the Bessel
function index $m=0$. However, we designate one group of modes as TM and another as TE, even for $m \ne 0$, simply
on the basis of the asymptotic limits attained by them. Notice that both asymptotic limits are correctly dictated
by Eqs. (3.53) and (3.54) in Ref. 115. What is noteworthy here is that the system supports the simultaneous
existence of TM and TE modes. It is also interesting to notice that the resonance frequency $\omega_0$ is not seen
to play any of its characteristic role in the spectrum (see in what follows).

%fig. 10
\begin{figure}[htbp]
\includegraphics*[width=7.5cm,height=8cm]{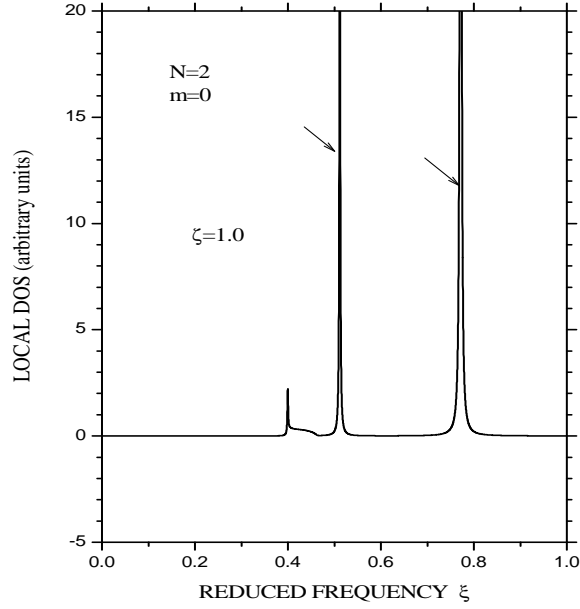}
%%%%%%%%%%%
\caption{Local density of states for the system discussed in Fig. 9 and for $m=0$ and $\zeta=1.0$. The rest of
the parameters used are the same as in Fig. 9. The arrows in the panel indicate the peaks at $\xi=0.5119$ and  $\xi=0.7718$. (After Kushwaha and Djafari-Rouhani, Ref. 115).}
\label{fig10}
\end{figure}

Figure 10 shows the local density of states for the system discussed in Fig. 9 for $m=0$ and $\zeta=1.0$. This
value of the propagation vector lies in the non-radiative region which allows the pure confined modes. The rest
of the parameters are the same as in Fig. 9. Both sharp peaks occurring at $\xi=0.5119$ and $\xi=0.7718$
reproduce exactly the respective TE and TM modes existing at $\zeta=1.0$ in Fig. 9. The short peak and the
related noisy part in the immediate vicinity of the resonance frequency $\omega_0$ has no physical significance
and will show its signature almost everywhere in the computation of local as well as total density of states.
It has been found that similar calculation of LDOS at any value of the propagation vector correctly reproduces
both modes in spectrum discussed in Fig. 9.

%fig. 11
\begin{figure}[htbp]
\includegraphics*[width=7.5cm,height=8cm]{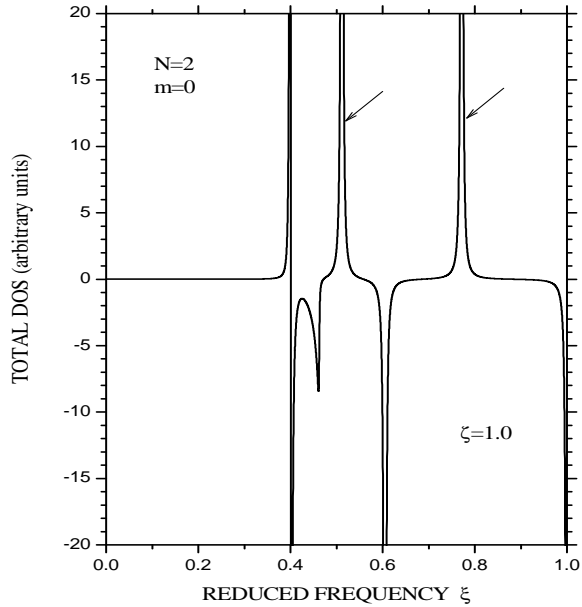}
%%%%%%%%%%%
\caption{Total density of states for the system discussed in Fig. 9 and for $m=0$ and $\zeta=1.0$. The rest of
the parameters used are the same as in Fig. 9. Both negative peaks are characteristic of the
resonance frequency $\omega_0$ and other characteristic frequency $\omega_c$ in the system and bear no physical significance. (After Kushwaha and Djafari-Rouhani, Ref. 115).}
\label{fig11}
\end{figure}

Figure 11 depicts the total density of states for the dielectric (vacuum) cable embedded in a dispersive
metamaterial background discussed in Fig. 9 for $m=0$ and $\zeta=1.0$. For $m=0$, the otherwise coupled
modes are decoupled as TM and TE. The $\zeta=1.0$ indicates the positions of the TE and TM modes lying at
$\xi=0.5119$ and $\xi=0.7718$ (see Fig. 9). Both of these positions of the respective modes are exactly
reproduced by the peaks marked by arrows in the total density of states here. A kind of resonant behavior
at $\xi\simeq 0.4$ and the negative peak at $\xi\simeq 0.6$ are characteristic of the critical frequencies
$\omega_0$ and $\omega_c$ involved in the problem and we do not consider them of any importance and/or
physical significance. Similar behavior at these frequencies will be seen in the later examples as well.

%fig. 12
\begin{figure}[htbp]
\includegraphics*[width=7.5cm,height=8cm]{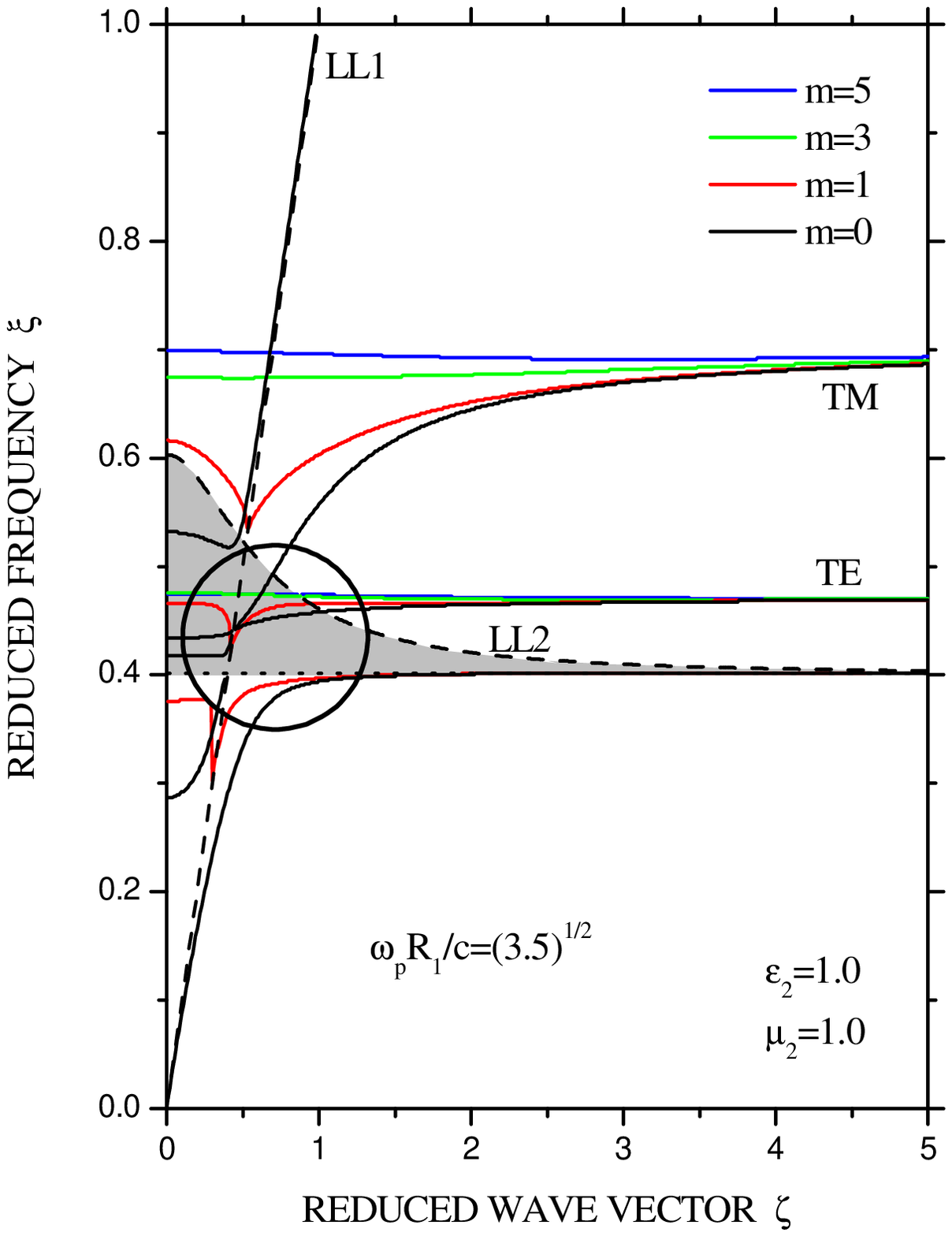}
%%%%%%%%%%%
\caption{(Color online) Plasmon dispersion for a metamaterial cable in a dielectric (vacuum) background for
different values of index $m=0$, 1, 3, 5. The dimensionless plasma frequency used in the computation is
specified by $\omega_pR_1/c=\sqrt{3.5}$. Dashed line and curve marked as LL1 and LL2 refer, respectively, to
the light lines in the vacuum and the metamaterial. The horizontal dotted line stands for the characteristic
resonance frequency ($\omega_0$) in the metamaterial. The shaded area represents the region within which both $\epsilon(\omega)$ and $\mu(\omega)$ are negative and disallows the existence of the confined modes.
(After Kushwaha and Djafari-Rouhani, Ref. 115).}
\label{fig12}
\end{figure}

Figure 12 illustrates the plasmon dispersion for a metamaterial cable in a dielectric (vacuum) background for
different values of index $m=0$, 1, 3, 5. The results are plotted in terms of the dimensionless frequency
$\xi$ and the dimensionless propagation vector $\zeta$. The dimensionless plasma frequency used in the
computation is specified by $\omega_pR_1/c=\sqrt{3.5}$. The dashed line and curve marked as LL1 and LL2 refer, respectively, to the light lines in the vacuum and the metamaterial. The horizontal dotted line stands for the characteristic resonance frequency ($\omega_0$) in the metamaterial. The shaded area represents the region
where both $\epsilon(\omega)$ and $\mu(\omega)$ are negative and disallows the existence of confined modes. We
designate the two groups of modes as TM and TE with the same notion as discussed in Fig. 9. One can see it
clearly that the resonance frequency $\omega_0$ does play a crucial role in this case. For instance, the big
hollow circle encloses the $m=0$ and the $m=1$ TM modes split due to the resonance frequency in the problem. We
call this splitting occurring between respective TM modes since we can see it happening just by plotting the
$m=0$ modes. If it were not for the resonance frequency the split $m=0$ mode would start from zero (just as
here) and propagate smoothly to approach the asymptotic limit without any splitting and $m=1$ mode would emerge
from a nonzero frequency somewhere in the radiative region. This is to stress that such resonance splitting
takes place only for the TM modes and not for the TE modes. The latter always start from above the resonance
frequency. It is noticeable that the split modes below $\omega_0$ later become asymptotic to $\omega_0$.  The
TM modes' splitting behavior will become more transparent in the later examples on double-interface systems (see,
for example, Figs. 15 and 18).

%fig. 13
\begin{figure}[htbp]
\includegraphics*[width=7.5cm,height=8cm]{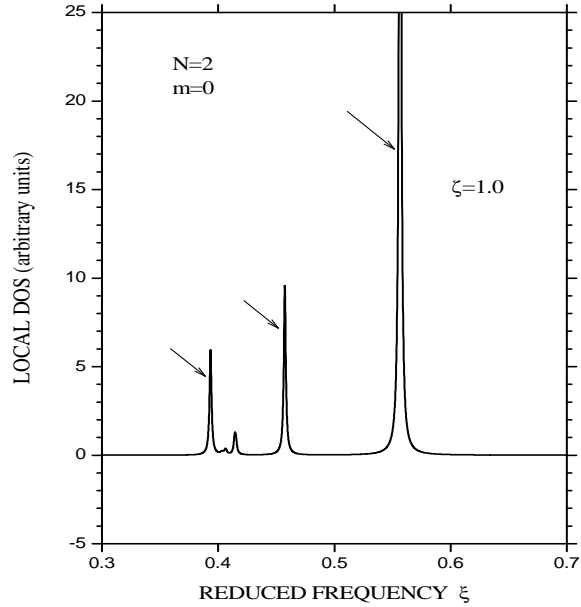}
%%%%%%%%%%%
\caption{Local density of states for the system discussed in Fig. 12 and for $m=0$ and $\zeta=1.0$. The arrows
in the panel indicate the peaks at $\xi=0.3947$, $\xi=0.4581$, and  $\xi=0.5577$. The rest of the parameters
used are the same as in Fig. 12. (After Kushwaha and Djafari-Rouhani, Ref. 115).}
\label{fig13}
\end{figure}

Figure 13 shows the local density of states for the system discussed in Fig. 12 for $m=0$ and $\zeta=1.0$. This
value of the propagation vector lies in the non-radiative region which allows the pure confined modes. The rest
of the parameters are the same as in Fig. 12. All the three sharp peaks lying at $\xi=0.3947$, $\xi=0.4581$, and  $\xi=0.5577$ reproduce exactly the respective TE and TM modes existing at $\zeta=1.0$ in Fig. 12. The short peak
and the related noise in the immediate vicinity of the resonance frequency $\omega_0$ has no physical significance
in the problem. It has been found that similar calculation of LDOS at any value of the propagation vector and for
any given $m$ correctly reproduces all the modes in spectrum discussed in Fig. 12. It should be pointed out that
while the LDOS (and/or TDOS, for that matter) can and do reproduce lowest (TM) split mode at the higher
propagation vector where this mode has already become asymptotic to and merged with $\omega_0$, sometimes it
becomes extremely difficult to discern such a peak inside the band of noise existing in the immediate vicinity of $\omega_0$.

%fig. 14
\begin{figure}[htbp]
\includegraphics*[width=7.5cm,height=8cm]{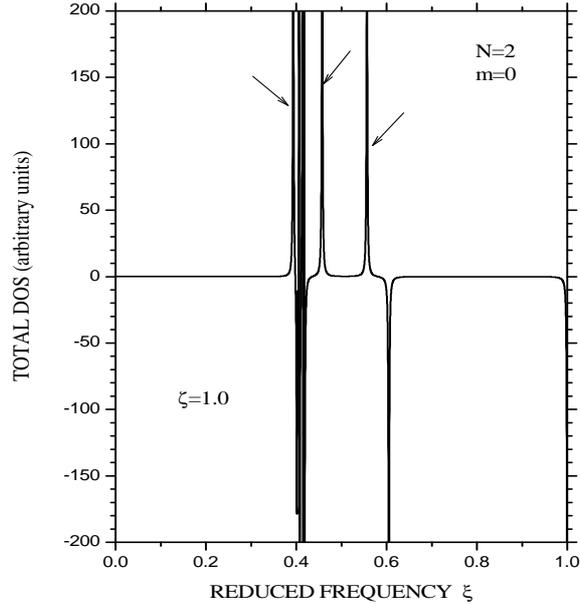}
%%%%%%%%%%%
\caption{Total density of states for the system discussed in Fig. 12 and for $m=0$ and $\zeta=1.0$. The arrows
in the panel indicate the peaks at $\xi=0.3947$, $\xi=0.4581$, and  $\xi=0.5577$. The rest of
the parameters used are the same as in Fig. 12. (After Kushwaha and Djafari-Rouhani, Ref. 115).}
\label{fig14}
\end{figure}

Figure 14 depicts the total density of states for the metamaterial cable embedded in a dispersive dielectric
(vacuum) background discussed in Fig. 12 for $m=0$ and $\zeta=1.0$. For $m0$, the otherwise coupled modes are
decoupled as TM and TE. The $\zeta=1.0$ specifies the positions of the decoupled modes lying at $\xi=0.3947$, $\xi=0.4581$, and  $\xi=0.5577$. All of these positions of the respective modes are correctly reproduced by
the peaks marked by the arrows in the total density of states here. The noisy band of states at $\omega_0$ and
the negative peak at $\omega_c$ are a consequence of these critical frequencies involved in the problem but
they carry no interesting information and bear no physical significance. Scanning the whole range of propagation
vector reveals that the TDOS reproduces all the modes in Fig. 12 very accurately. The only exception to this is
the radiative modes (towards the left of the light line) in Fig. 12, which do not show a good correspondence
with the resonance peaks in the (local or total) DOS. This is not surprising, however, given the distinct ways
of searching the zeros of the complex transcendental function in the radiative and non-radiative regions. We
did not intend to pay much attention to the small radiative region simply because, as we all know, this region
is of almost no practical interest.

\subsection{Double-interface systems}

%fig. 15
\begin{figure}[htbp]
\includegraphics*[width=7.5cm,height=8cm]{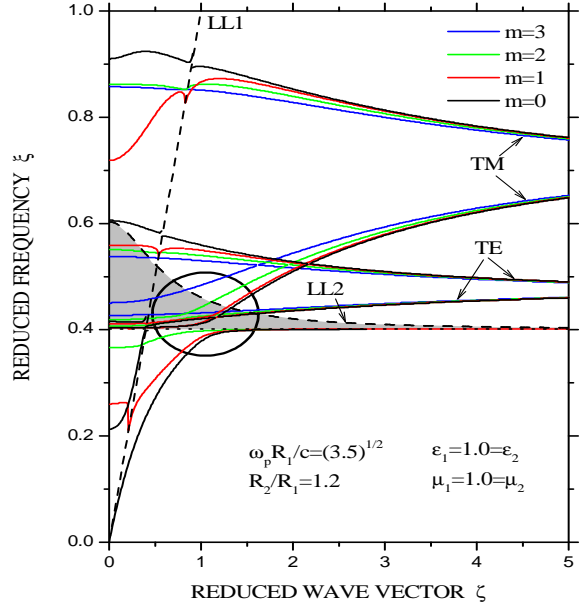}
%%%%%%%%%%%
\caption{(Color online) Plasmon dispersion for a metamaterial shell sandwiched between two identical dielectrics
(vacuum) for different values of index $m=0$, 1, 2, 3. The dimensionless plasma frequency used here is specified
by $\omega_pR_1/c=\sqrt{3.5}$ and the radii ratio $R_2/R_1=1.2$. Dashed line and curve marked as LL1 and LL2
refer, respectively, to the light lines in the vacuum and the metamaterial. The horizontal dotted line stands for
the characteristic resonance frequency ($\omega_0$) in the metamaterial. The shaded area represents the region
within which both $\epsilon(\omega)$ and $\mu(\omega)$ are negative and forbids the existence of the confined
modes. The parameters used in the computation are as listed in the picture.
(After Kushwaha and Djafari-Rouhani, Ref. 115).}
\label{fig15}
\end{figure}

Figure 15 illustrates the plasmon dispersion for a metamaterial shell sandwiched between two identical dielectrics (vacuum) for different values of index $m=0$, 1, 2, and 3. The plots are rendered in terms of the dimensionless
frequency $\xi$ and the propagation vector $\zeta$. The other important parameters used in the computation are the normalized plasma frequency $\omega_pR_1/c=\sqrt{3.5}$ and the ratio $R_2/R_1=1.2$. The dashed line and the curve
marked as LL1 and LL2 refer, respectively, to the light lines in the bounding media (vacuum) and the metamaterial
shell. The horizontal dotted line refers to the characteristic resonance frequency ($\omega_0$) in the problem.
The shaded area refers to the region where $\epsilon(\omega) < 0$ and $\mu(\omega) < 0$ and prohibits the existence
of the truly confined modes. Since there are two interfaces involved in the resultant structure we have a pair of
modes for each of the TM and TE modes in the system. The lower and upper group of modes together attain the same asymptotic limit characteristic of the TM or the TE modes at large wave vectors. The resonance frequency allows
the splitting of the $m=0$, $m=1$, and $m=2$ TM modes at $\omega_0$. The full circle encloses and shows such a
resonance splitting occurring between the respective modes in a very clear way at $\zeta \simeq 1.0$. The scheme of assigning the modes a TM or a TE character is the same as discussed before (see discussion of Fig. 9). Eqs. (3.53)
and (3.54) in Ref. 115 are seen to dictate the correct asymptotic limits attained both by TM and TE modes. Some
abruptness (sharp or blunt) observed by a given mode at the light line is a general tendency usually seen when a
mode crosses the junction between the two media.

%fig. 16
\begin{figure}[htbp]
\includegraphics*[width=7.5cm,height=8cm]{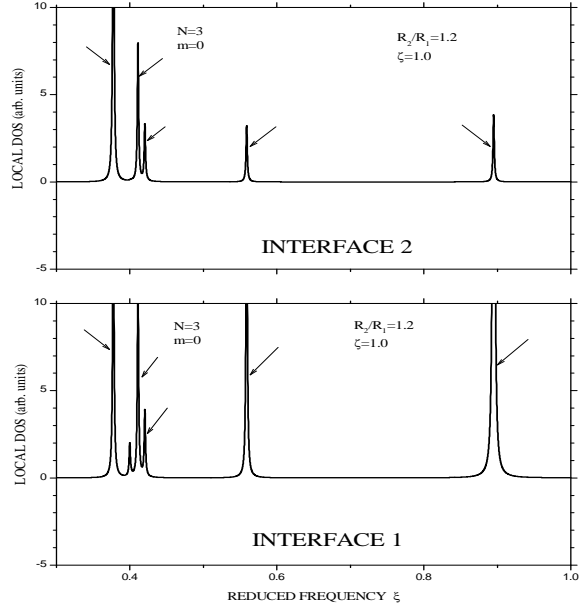}
%%%%%%%%%%%
\caption{Local density of states at the interface $R_1$ ($R_2$) in the lower (upper) panel for $m=0$ and
$\zeta=1.0$ for the system discussed in Fig. 15. We call attention to the DOS resonance peaks, indicated
by the arrows, corresponding to the five modes in total at $\zeta=1.0$ in Fig. 15. The interface 1 (2)
refers to the one specified by $R_1$ ($R_2$). The rest of the parameters used are the same as in Fig. 15.
(After Kushwaha and Djafari-Rouhani, Ref. 115).}
\label{fig16}
\end{figure}

Figure 16 shows the local density of states for the two-interface system discussed in Fig. 15 for $m=0$ and
$\zeta=1.0$ for the interface 1 (2) in the lower (upper) panel. This value of $\zeta$ specifies five
propagating modes in total in Fig. 15: the lowest split (TM) mode below $\omega_0$, lower (split) TM mode
and lower TE mode within the shaded region, upper TE mode, and the uppermost TM mode lying, respectively,
at $\xi=0.3774$, $\xi=0.4109$, $\xi=0.4202$, $\xi=0.5588$, and $\xi=0.8951$. In the lower panel, the five
resonance peaks (indicated by arrows) observed in the local density of states stand exactly at these
frequencies. This implies reasonably a very good correspondence between the (dispersion) spectrum and the
LDOS at interface $R_1$. The (unmarked) second lowest peak (counting from the lowest frequency) stands at
the resonance frequency $\omega_0$ and is not considered to be a bonafide peak in the LDOS. Coming to the
upper panel, we again observe five well-defined resonance peaks lying exactly at the aforementioned
frequencies. That means that both interfaces share all the five resonances in the LDOS, albeit with a
difference of magnitude. This also implies that the two interfaces pose different preferences, and that
makes sense here because of the asymmetric nature of the configuration. In other words, the two interfaces
seem to be more sensitive to the geometry and less to the materials in the supporting media. That is to say
that the situation is altogether different from a planar geometry with, for example, a thin metallic or
semiconducting film symmetrically bounded by two identical dielectrics.

%fig. 17
\begin{figure}[htbp]
\includegraphics*[width=7.5cm,height=8cm]{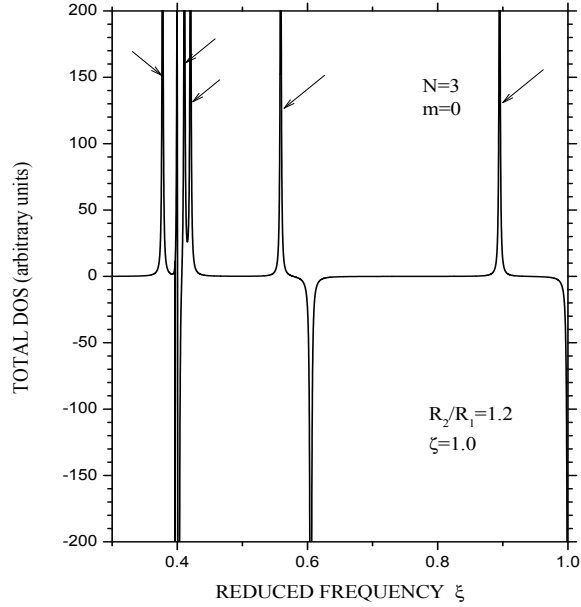}
%%%%%%%%%%%
\caption{Total density of states for $m=0$ and $\zeta=1.0$ for the system discussed in Fig. 15. We call attention
to the DOS resonance peaks, indicated by the arrows, corresponding to the five modes in total at $\zeta=1.0$ in
Fig. 15. The parameters used are the same as in Fig. 15. (After Kushwaha and Djafari-Rouhani, Ref. 115).}
\label{fig17}
\end{figure}

Figure 17 depicts the total density of states for the two-interface system discussed in Fig. 15 for $m=0$ and
$\zeta=1.0$. These values of $m$ and $\zeta$ define five propagating modes in total in Fig. 15, covering both
TM and TE modes, and lying at $\xi=0.3774$, $\xi=0.4109$, $\xi=0.4202$, $\xi=0.5588$, and $\xi=0.8951$. The
five resonance peaks (indicated by arrows) observed in the total density of states are seen to substantiate
these frequencies very accurately. We have seen that the similar computation of TDOS at any other value of
$\zeta$ reproduces all the corresponding modes in the spectrum very correctly. The only exception to this is
the radiative region (toward the left of the light line) where the correspondence between the DOS (local or
total) and the spectrum is no so good. This is again understandable in the view of the facts stated above (see
the discussion of Fig. 14). A pile up of the states at and in the vicinity of $\omega_0$ and a negative peak
at $\omega_c$ are clearly a consequence of the presence of such critical (resonance) frequencies in the problem
and we do not consider them to be of any physical significance.

%fig. 18
\begin{figure}[htbp]
\includegraphics*[width=7.5cm,height=8cm]{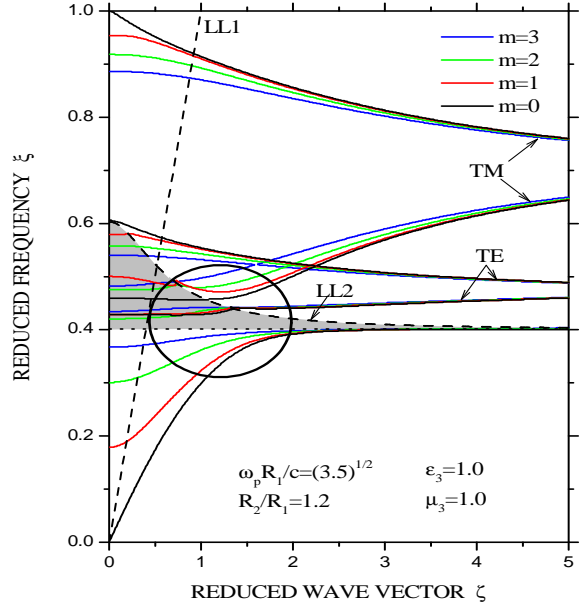}
%%%%%%%%%%%
\caption{(Color online) Plasmon dispersion for a dielectric (vacuum) shell sandwiched between two identical
metamaterials for different values of index $m=0$, 1, 2, 3. The dimensionless plasma frequency used here is
specified by $\omega_pR_1/c=\sqrt{3.5}$ and the radii ratio $R_2/R_1=1.2$. Dashed line and curve marked as
LL1 and LL2 refer, respectively, to the light lines in the vacuum and the metamaterial. The horizontal
dotted line stands for the characteristic resonance frequency ($\omega_0$) in the metamaterial. The shaded
area represents the region within which both $\epsilon(\omega)$ and $\mu(\omega)$ are negative and proscribes
the existence of the confined modes. The parameters used in the computation are as listed in the picture.
(After Kushwaha and Djafari-Rouhani, Ref. 115). }
\label{fig18}
\end{figure}

Figure 18 illustrates the plasmon dispersion for a dielectric (vacuum) shell sandwiched between two identical metamaterials for different values of index $m=0$, 1, 2, and 3. The plots are rendered in terms of the
dimensionless frequency $\xi$ and the propagation vector $\zeta$. The other important parameters used in the
problem are the normalized plasma frequency $\omega_pR_1/c=\sqrt{3.5}$ and the ratio of the radii $R_2/R_1=1.2$.
The dashed line and the curve marked as LL1 and LL2 stand, respectively, for the light lines in the dielectric
(vacuum) and the bounding metamaterials. The horizontal dotted line refers to the characteristic resonance
frequency ($\omega_0$) in the problem. The shaded area refers to the region within which $\epsilon(\omega) < 0$
and $\mu(\omega) < 0$ and proscribes the existence of the truly confined modes. Again, since there are two
interfaces in the system we obtain two pairs of modes: one for the TM and the other for the TE modes. Their
asymptotic limits are governed by Eqs. (3.53) and (3.54) in Ref. 115. The presence of the resonance frequency
$\omega_0$ gives rise to the resonance splitting of all the lower group of the pair of TM modes for different
values of $m$. Although the lower group of the pair of TE modes cross in between the split TM modes (inside the
shaded region), the resonance splitting is clearly pronounced between the corresponding TM modes. This is shown
by the big hollow circle encompassing all the respective split TM modes in the region. Again, the scheme of
assigning the modes a TM or a TE character is the same as discussed before. Notice that the abruptness observed
by the modes while crossing the light line is relatively smoother than that seen in the other cases (cf. Figs.
9, 12, and 15). It is interesting to remark that all the illustrative examples on the plasmon spectrum presented
here reaffirm that the dispersive metamaterial components in the composite enable the structure to support the simultaneous existence of the TM and the TE modes. This effect is solely attributed to the negative-index
metamaterials and is otherwise impossible.

%fig. 19
\begin{figure}[htbp]
\includegraphics*[width=7.5cm,height=8cm]{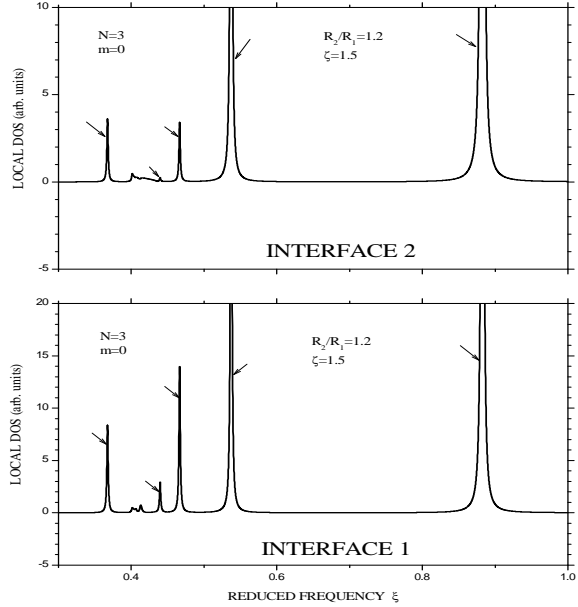}
%%%%%%%%%%%
\caption{Local density of states at the interface $R_1$ ($R_2$) in the lower (upper) panel for $m=0$ and
$\zeta=1.5$ for the system discussed in Fig. 18. We call attention to the DOS resonance peaks, indicated
by the arrows, corresponding to the five modes in total at $\zeta=1.5$ in Fig. 18. The interface 1 (2)
refers to the one specified by $R_1$ ($R_2$). The rest of the parameters used are the same as in Fig. 18.
(After Kushwaha and Djafari-Rouhani, Ref. 115).}
\label{fig19}
\end{figure}

Figure 19 shows the local density of states for the two-interface system discussed in Fig. 18 for $m=0$ and
$\zeta=1.5$ for the interface 1 (2) in the lower (upper) panel. This value of $\zeta$ characterizes five
propagating modes in total in Fig. 11: the lowest split (TM) mode below $\omega_0$, lower TE mode, lower
(split) TM mode, upper TE mode, and the uppermost TM mode lying, respectively, at $\xi=0.3673$, $\xi=0.4396$, $\xi=0.4665$, $\xi=0.5373$, and $\xi=0.8825$. In the lower panel, the five resonance peaks (indicated by
arrows) observed in the local density of states stand exactly at these frequencies. This implies considerably
a very good correspondence between the (dispersion) spectrum and the LDOS at interface $R_1$. The small
(unmarked) noisy peaks occurring in the vicinity of the resonance frequency $\omega_0$ are not considered to
be a bonafide peaks in the LDOS. In the upper panel, we plot the LDOS for the interface 2 for the same
parameters as considered for interface 1 in the lower panel. We observe five well-defined resonance peaks
lying exactly at the aforementioned frequencies. That means that both interfaces share all the five resonances
in the LDOS, of course with a difference of magnitude. As to the second small resonance peak in this panel,
we think that interface 2 only slightly feels this resonance. Other resonance peaks in this panel are almost
comparable to those in the lower panel. The rest of the remarks made with respect to Fig. 16 are also valid
here.

%fig. 20
\begin{figure}[htbp]
\includegraphics*[width=7.5cm,height=8cm]{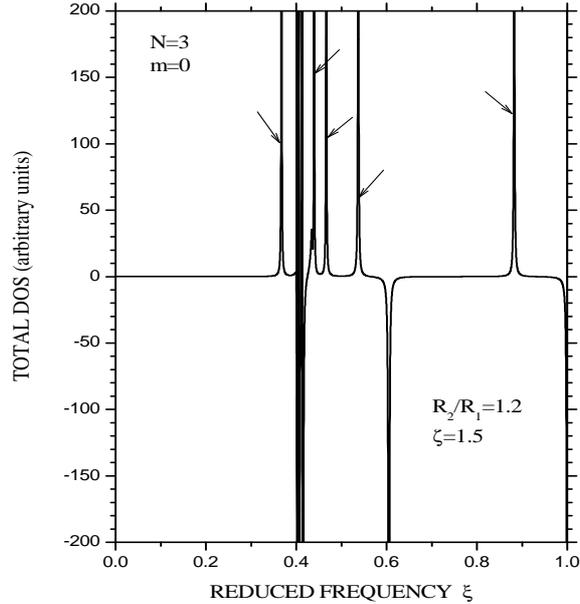}
%%%%%%%%%%%
\caption{Total density of states for $m=0$ and $\zeta=1.5$ for the system discussed in Fig. 18. We call
attention to the DOS resonance peaks, indicated by the arrows, corresponding to the five modes in total
at $\zeta=1.5$ in Fig. 18. The parameters used are the same as in Fig. 18. The DOS are shown in arbitrary
units throughout. (After Kushwaha and Djafari-Rouhani, Ref. 115).}
\label{fig20}
\end{figure}

Figure 20 depicts the total density of states for the same two-interface system as investigated in Figs. 18
and 19 for $m=0$ and $\zeta=1.5$. Such values of $m$ and $\zeta$ characterize five propagating modes in total
in Fig. 18, covering both TM and TE modes, and lying at $\xi=0.3673$, $\xi=0.4396$, $\xi=0.4665$, $\xi=0.5373$,
and $\xi=0.8825$. We observe that there are five well-defined resonance peaks in the total density of states
standing exactly at the aforementioned frequencies. This leads us to infer that there is a very good
correspondence between the (dispersion) spectrum and the TDOS. It has been observed that scanning other values
of the propagation vector $\zeta$ for computing the total density of states yields same degree of correspondence
with the spectrum.  What is more interesting in this case is the fact that the computation of TDOS (as well as
LDOS) provides a much better correspondence with the modes in the spectrum even in the radiative region (toward
the left of the light line) than in the previous cases. This is attributed to a relatively smoother propagation
of the modes in the radiative region in the present case of a dielectric shell bounded by (identical)
metamaterials. Just as before, we do not give much importance to the pile up of the states near the resonance
frequency $\omega_0$ and the negative peak at $\omega_c$. While we consider their occurrence as natural, they
do not bear any physical significance to the problem whatsoever.

\subsection{Multicoaxial cable systems}

This section is devoted to investigate the plasmon excitations in a neat and clean system comprised of
multicoaxial negative-index metamaterial cables, schematically shown in Fig. 21. The formalism of the
problem is a straightforward generalization of the theory presented in Sec. III of Ref. 115. This is
systematically illustrated in Fig. 22, with all the necessary details. Since these are, to our knowledge,
the first results on such a complex system of multicoaxial metamaterial cables (MCMC), we adhere to the
backbone simplicity and think that the complexities, such as the choice of different metamaterials,
different (conventional) dielectrics as spacers, different (irregular) radii, and geometrical defects
would (and should) come later and hence are deferred to a future publication. As such, our system is
considered to be made up of the (same) RHM and the (same) LHM in alternate shells starting from the
innermost cable of radius $R_1$ and fix (the total number of media, including the outermost semi-infinite
medium) $n=15$.

%fig. 21
\begin{figure}[htbp]
\includegraphics*[width=7cm,height=6cm]{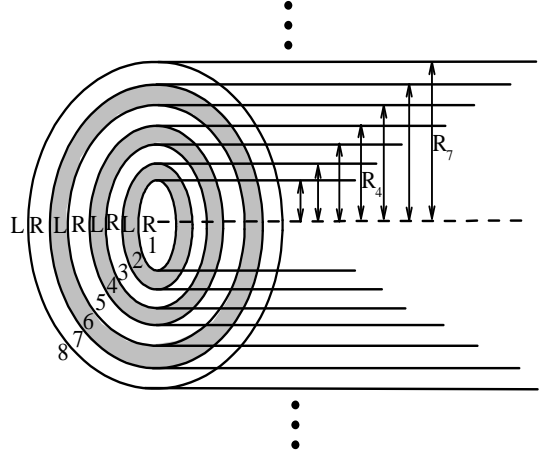}
%%%%%%%%%%%
\caption{Schematics of the multi-coaxial cables: the side view showing the alignments of the cylindrical
cables of circular cross-sections of radii $R_{j+1} > R_j$, and $n$ media with $n-1$ interfaces. The
innermost circle marked as 1 refers to the innermost cable of radius $R_1$ enclosed by the consecutive
($n-1$) shells assumed to be numbered as 2, 3, .... (n-2), (n-1) and cladded by an outermost
semi-infinite medium $n$. Our exact general theory schematically outlined in Fig. 15 allows one to
consider the resultant system to be made up of negative-index (dispersive or nondispersive)
metamaterials interlaced with conventional dielectrics, metals, or semiconductors. (After Kushwaha and
Djafari-Rouhani, Ref. 116).}
\label{fig21}
\end{figure}

%%%%%%%%%%%%%%%%%%%%%%%%%%%%%

%fig. 22
\begin{figure}[htbp]
\includegraphics*[width=7.5cm,height=7cm]{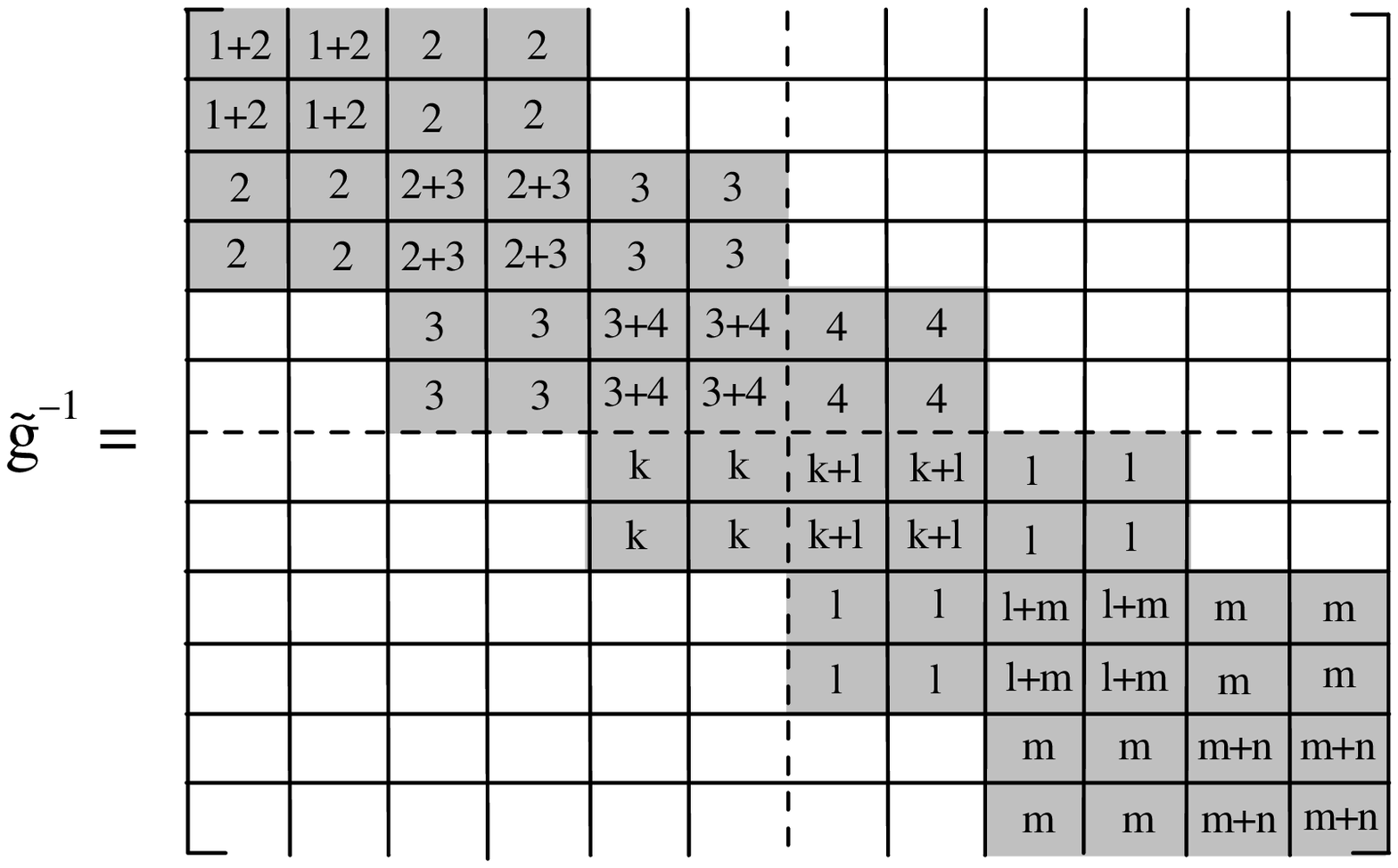}
%%%%%%%%%%%
\caption{A graphic representation of the complete formalism for the total inverse response function
$\tilde{g}^{-1}(...)$ for the resultant system shown in a desired compact form. Here $n$ refers to the
total number of media comprising the MCMC system; with $m=n-1$ as the number of interfaces, $l=n-2$,
and $k=n-3$ ..... etc. {\it We call attention to the fact that this is a $2(n-1)\times 2(n-1)$ matrix,
with all the elements outside the shaded regions being zero}. Now `1' refers to the first perturbation
specified by Eq. (3.8), `n' stands for the second perturbation specified by Eq. (3.15), and `2', `3',
`4', ....., `(n-2)', `(n-1)' correspond to the third perturbation specified by Eq. (3.23) for the
respective shells [115]. We would like to stress that our formalism is {\it not} a perturbative scheme,
albeit we use the term `perturbation' --- the term `perturbation', in fact, implies to the step-wise
operation concerned with the problem. It is also noteworthy that this theoretical framework knows no
bound with respect to the number of media involved in the system and/or their material characteristics.
The plasma modes of the system are defined by $det [\tilde{g}^{-1}(...)]=0$. (After Kushwaha and
Djafari-Rouhani, Ref. 116).}
\label{fig22}
\end{figure}

Figure 23 illustrates the surface plasmon dispersion for a perfect multicoaxial cable system made up of a
dispersive negative-index metamaterials interlaced with conventional dielectrics (assumed to be vacuum)
for $n=15$ and (the integer order of the Bessel function) $m=0$. The plots are rendered in terms of the
dimensionless frequency $\xi=\omega/\omega_p$ and the propagation vector $\zeta=c k/\omega_p$. The dashed
line and curve marked as LL1 and LL2 refer, respectively, to the light lines in the vacuum and the
metamaterial. The shaded area represents the region within which both $\epsilon(\omega)$ and $\mu(\omega)$
are negative and disallows the existence of truly confined modes. The thick dark band of frequencies piled
up near the resonance frequency $\omega_0=0.4 \omega_p$ is not unexpected. Since there are fourteen
interfaces in the system, we logically expect fourteen branches each for the TM and TE modes in the system.
As we see, this is exactly the case, except for the fact that the lower group of seven TM branches (which
start from zero) have observed a resonance splitting due the resonance frequency $\omega_0$ in the problem.
The latter branches quickly become asymptotic to $\omega_0$. All the TM and TE confined modes above
$\omega_0$ have their well-defined asymptotic limits exactly dictated, respectively, by Eqs. (3.53) and
(3.54) [115]. A word of warning about the simultaneous existence of TM and TE modes: if we search the
zeros of the determinant (see Fig. 22), as it is required, for {\em any} value of $n$ and $m$, we always
obtain the simultaneous existence of all the TM and TE modes along with the resonance splittings as stated
above. This is a {\em rule} as long as $m \ne 0$. The only exception to this is the case of $m=0$ (and very
small value of $n$). For instance, for $n=3$ and $m=0$, one has a $4 \times 4$ determinant and it is possible
to separate analytically the TM and TE modes. [We recall the well-known facts from the electrodynamics: the
electrostatics claim ownership of the p-polarized (TM) fields and the magnetostatics claim the s-polarized
(TE) fields.] However, even for these values of $n$ and $m$, if we search the zeros of the full determinant,
without analytically decoupling the modes, we obtain both TM and TE modes together.

%fig. 23
\begin{figure}[htbp]
\includegraphics*[width=7.5cm,height=8cm]{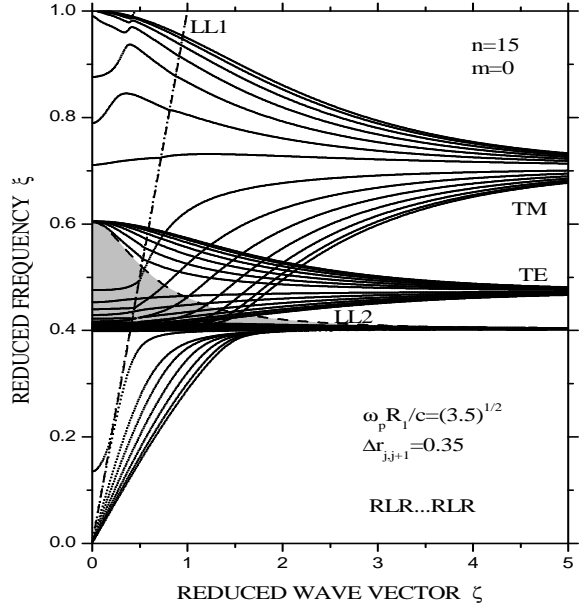}
%%%%%%%%%%%
\caption{Plasmon dispersion for a perfect multicoaxial cable system made up of a dispersive negative-index
metamaterial interlaced with conventional dielectrics (assumed to be vacuum) for $N=15$ and $m=0$. The
dimensionless plasma frequency used in the computation is specified by $\omega_pR_1/c=\sqrt{3.5}$ and the
dimensionless thicknesses of the shells are defined by $\Delta r_{j,j+1}=0.35$. Dashed line and
curve marked as LL1 and LL2 refer, respectively, to the light lines in the vacuum and the metamaterial.
The (dark) thick band of frequencies is piled up at the characteristic resonance frequency ($\omega_0$) in
the metamaterial. We call attention to the resonance splitting of the lower TM confined modes due to the
resonance frequency ($\omega_0$) in the problem. The shaded area represents the region within which both $\epsilon(\omega)$ and $\mu(\omega)$ are negative and disallows the existence of truly confined modes. The
system as a whole is represented by RLR...RLR design. (After Kushwaha and Djafari-Rouhani, Ref. 116).}
\label{fig23}
\end{figure}

Figure 24 shows the local density of states (LDOS) as a function of reduced frequency $\xi$ for the multicoaxial
cable system discussed in Fig. 23, for the propagation vector $\zeta=1.0$. The rest of the parameters used are
the same as those in Fig. 23. Notice that each of these interfaces is seen to share most of the peaks supposed
to exist and reproduce most of the discernible modes at $\zeta =1.0$ (in Fig. 23). Of course, one has to take
into consideration the degeneracy and the hodgepodge that persists near the resonance frequency $\xi=0.4$ in
Fig. 23. Let us, for instance, look at top panel: the highest, second highest, third highest, fourth highest,
fifth highest, sixth highest, seventh highest, and eighth highest peaks lie, respectively,
at $\xi=0.9490$, 0.9428, 0.9299, 0.9061, 0.8667, 0.8046, 0.7298, and 0.6232. The highest peak (at $\xi=0.9490$)
remains indiscernible at this scale.  Similarly, the lowest, second lowest, third lowest, fourth lowest, fifth
lowest, sixth lowest, and seventh lowest peaks (below the resonance frequency) are seen to lie, respectively, at $\xi=0.2877$, 0.2999, 0.3197, 0.3452, 0.3663, 0.3763, and 0.3967. All these peak positions exactly substantiate
the modes at $\zeta=1.0$ in the spectrum in Fig. 23. One has to notice that as the name {\em local} DOS suggests,
every interface has its own choice (with respect to the geometry and/or the material parameters) and there does
not seem to be a rule that may dictate the modes' counting. This is, in a sense, different from the total DOS
where one obtains exactly the same number of peaks as the modes in the spectrum for a given value of $\zeta$. It
should be pointed out that the shorter-wavelength modes do not interact much with the neighboring ones and remain spatially confined to the immediate vicinity of the respective interfaces. Such strongly localized modes are thus
easier to be observed in the experiments than their longer-wavelength counterparts.

%fig. 24
\begin{figure}[htbp]
\includegraphics*[width=7.5cm,height=8cm]{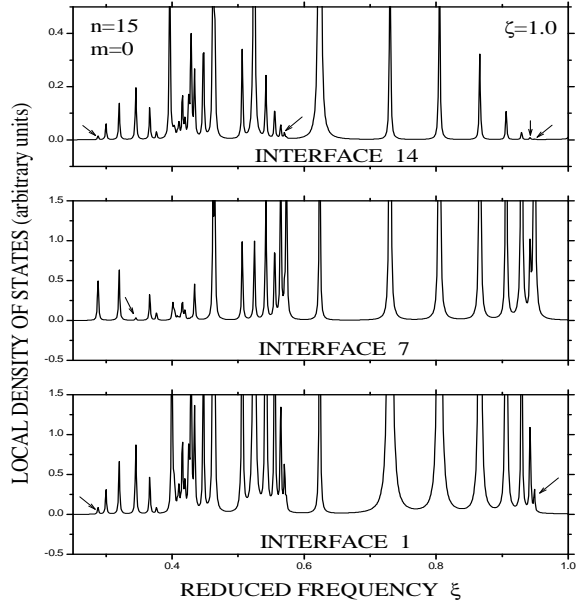}
%%%%%%%%%%%
\caption{Local density of states for the system discussed in Fig. 23 and for $n=15$, (the integer order of
the Bessel function) $m=0$, and the reduced propagation vector $\zeta=1.0$. The bottom, middle, and top
panel refer, respectively, to LDOS at interface 1, interface 7, and interface 14 in the system. The rest of
the parameters used are the same as in Fig. 23. The arrows in the panels indicate the relatively smaller
(in height) peaks. (After Kushwaha and Djafari-Rouhani, Ref. 116).}
\label{fig24}
\end{figure}

%%%%%%%%%%%%%%%%%%%%%%%%%%%%%

%fig. 25
\begin{figure}[htbp]
\includegraphics*[width=7.5cm,height=8cm]{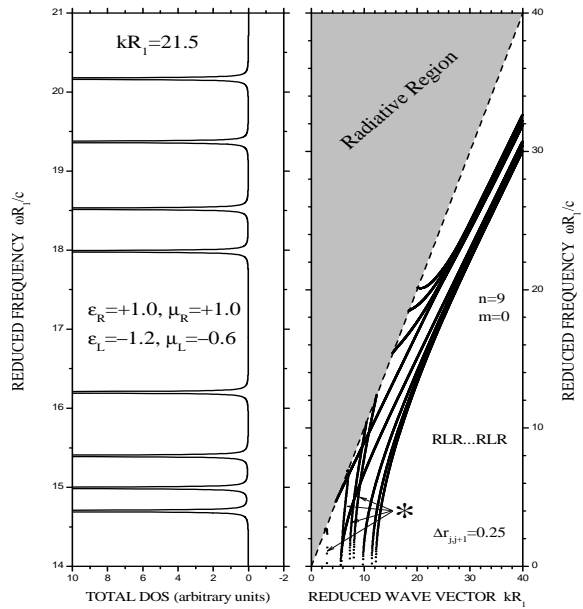}
%%%%%%%%%%%
\caption{Right panel: Plasmon dispersion for a perfect MCMC cable system made up of a non-dispersive
negative-index metamaterial interlaced with conventional dielectrics (assumed to be vacuum) for $n=9$
and (the integer order of the Bessel function) $m=0$. There are well-defined $n-1$ TM modes in the
system: upper four starting from the light-line and the lower four from the nonzero propagation vector.
The dimensionless thicknesses of the shells are defined by $\Delta r_{j,j+1}=0.25$. The shaded region
stands for the purely radiative modes (not shown). We call attention to the sharply downward modes,
indicated by arrows, which are the ill-behaved TE modes in the LWL. Left panel: the total density of
states as a function of reduced frequency $\omega R_1/c$ for the dimensionless wave vector $kR_1=21.5$.
(After Kushwaha and Djafari-Rouhani, Ref. 116).}
\label{fig25}
\end{figure}

Figure 25 depicts the surface plasmon dispersion (right panel) and total density of states (left panel) for
a multicoaxial cable system made up of a nondispersive negative-index metamaterials interlaced with
conventional dielectrics (assumed to be vacuum) for $n=9$ and (the integer order of the Bessel function)
$m=0$. The plots are rendered in terms of the dimensionless frequency $\omega R_1/c$ and the propagation
vector $k R_1$. The dimensionless thicknesses of the shells are defined by $\Delta r_{j,j+1}=0.25$. The
material parameters are listed inside the left panel. Right panel: The dashed line refers to the light line
in the vacuum. As expected, there are eight TM modes --- four of them starting from the nonzero propagation
vector $k$ and the other four emerge from the light line. The shaded area is the radiative region which
encompasses radiative modes (not shown) towards the left of the light line. We notice in passing that the
slope of these TM modes in the asymptotic limit is defined by $\omega/c k=0.7817$. An important issue remains
to be answered: Why do we obtain only TM modes up to the asymptotic limit? In order to answer this question,
one has to look carefully at the analytical diagnoses presented in Sec. III.G [115]. To be brief, the answer
lies in the fact that, for the material parameters chosen here, while Eq. (3.43) is fully satisfied,
Eq. (3.47) or (3.48) is not [see Ref. 115 for the analytical diagnosis]. The former condition justifies the
existence of the TM modes and the latter rules out the occurrence of TE modes. One remaining curiosity: What
do these (almost) vertical lines, hanging downwards from the light line, indicated by arrows refer to? The
succinct answer is that these are the ill-behaved TE modes which exist only in the long wavelength limit (LWL).
It is interesting to note that if we interchange the values of $\epsilon_L$ and $\mu_L$, (i.e., the parameters
that define the nondispersive LHM), we obtain well-behaved TE modes and the ill-behaved TM modes. The reason
is simple: the aforesaid conditions that govern the nature of the modes in the asymptotic limit are then
reversed. Left panel: The computation of the total density of states plotted as a function of reduced
frequency shows clearly eight peaks for the given value of the propagation vector $k R_1=21.5$. Starting from
the lowest frequency, we observe that these peaks lie
at $\omega R_1/c=14.69$, 14.99, 15.40, 16.20, 17.99, 18.53, 19.37, and 20.18. These peak positions exactly
substantiate the frequencies of the TM modes in the right panel at $k R_1=21.5$.

To conclude with, we estimate that if $\nu_p=\omega_p/2\pi=10$ GHz, the radius of the innermost cable is defined
as $R_1=8.93$ mm for the parameter $\omega_pR_1/c=1.87$ used for the dispersive negative-index metamaterials (see
Figs. 23 and 24). It is interesting to notice that this size scale is almost the same as the dimensions of the
sample (the lattice spacing $d=9.53$ mm and inner size of the tubes $a=6.96$ mm) used in the experiment in Ref.
111, which had verified the prediction of Pendry and coworkers [103] that, if textured on a subwavelength scale,
even perfect conductors can support the surface plasmon modes. The surface plasmon modes predicted here should
be observable in the inelastic electron scattering (EELS) or inelastic light (Raman) scattering experiments. The
EELS is already becoming known to be a powerful probe for studying the plasmon excitations in single- and
multi-wall carbon nanotubes.

\section{TIDBITS OF THE APPLICATIONS OF METAMATERIALS}

To this point, we hope to have become sufficiently familiar with what, how, and why about the metamaterials. Yet
it is interesting to add that metamaterials are periodic arrangements of individual elements made up of
conventional (microscopic) materials such as metals or plastics. Metamaterials gain their properties from their exactingly-designed structures rather than from their composition. Their precise shape, size, geometry,
orientation, and arrangement can affect the waves of interest, creating material properties which
are unattainable with conventional materials. The structural elements making up the metamaterials are chosen to
be of subwavelength sizes -- the features that are actually smaller than the wavelength of the waves they are
intended to affect. The primary goal of the research in metamaterials is to investigate the negative-index
materials (NIMs), which appear to have allowed the creation of superlenses with a spatial resolution beyond the
diffraction limit [24]. The research in metamaterials is interdisciplinary and involves fields such as
electromagnetics, electrical engineering, electronics, classic optics, material sciences, solid state physics, semiconductor sciences, and nanoscience. Potential applications of metamaterials are diverse and include areas
such as high-frequency communication, remote aerospace, sensor design and detection, smart solar power, public
safety, shielding structures from temblor, superlenses for high-gain antennas, just to name a few.

In ordinary materials – solid, liquid, or gas; transparent or opaque; conductor or insulator – the conventional
refractive index predominates. This means that $\epsilon$ and $\mu$ are both positive resulting in an ordinary
(positive) index of refraction. However, metamaterials have the capability to exhibit a state where both
$\epsilon$ and $\mu$ are negative, resulting in an extraordinary (negative) index of refraction. Depending upon
the type of the waves and the frequency range of interest, the designed metamaterials are becoming known and
being distinguished by the wide variety of qualifiers such as single-negative (SNG) metamaterials, which
include $\epsilon$-negative (ENG) media and $\mu$-negative (MNG) media, and negative-index metamaterials (NIMs),
which include metamaterials with both $\epsilon$ and $\mu$ negative and hence the name double-negative (DNG)
materials. Note that the nature encompasses materials with $\epsilon$ and $\mu$ both positive, which are
becoming known as double positive media (DPS).  Artificial materials have been fabricated which have DPS, ENG,
and MNG properties combined [124].

A further classification of metamaterials is based on the types of designed materials, which interact at
different frequencies such as microwave, gigahertz, terahertz, and later, optical frequencies. The
corresponding metamaterials are given the nicknames such as acoustic, elastic, seismic, photonic, plasmonic,
superlens, chiral [125-126], bi-isotropic and bi-anisotropic [127-129], GHz, and THz metamaterials. The
literature is witnessing a relatively greater attention being paid to the designing of the cloaking devices,
however. The real challenge of cloaking lies in devising a theoretical scheme for the optical properties of
the cloak and even more challenging is realizing those properties in a material. Transformation optics [130]
provides the analytical background and metamaterials provide the means of achieving the prescribed parameters.
The work on designing optical [72], acoustic [85], and plasmonic [98] cloakings continues to be escalating.
Even though a working, practical cloak is still far from the sight, the design and debate seem to be wildly
optimistic.

\section{THE CLOSING REMARKS}

To conclude with, we have gathered and reviewed the fundamental aspects related to the concepts, nomenclature, fabrication, and applications of the metamaterials. In particular, we have stressed upon the propagation
characteristics of plasmon polaritons and their density of states in the coaxial cables in the absence of an
applied magnetic field. The illustrative numerical examples in Sec. III follow the brief strategy of working
within the Green-function theory for the plasmon propagation in any planar, cylindrical, or spherical
geometries in Sec. II. We have sagaciously attempted to refer the interested reader to the derivations of the
general results and details of their analytical diagnoses within the theoretical framework given in Ref. 115.
As to the illustrative examples, we have also successfully attempted to substantiate our results on plasmon
dispersion through the computation of the local and total density of states. While we considered the effect of retardation, the absorption was neglected throughout, except for a small imaginary part needed to be added to
the frequency for the purpose of giving a width to the peaks in the DOS. We believe that the present
methodology for coaxial cables will also prove to be a powerful theoretical framework for studying, for example,
the intrasubband plasmons in the multi-walled carbon nanotubes.

An experimental observation of the radiative as well as non-radiative plasmons in such coaxial cables would
be of great interest. Such experiments could possibly involve the well known attenuated total reflection,
scattering of high energy electrons, or even Raman spectroscopy. The electron energy loss spectroscopy (EELS)
is already becoming known as a powerful technique for studying the electronic structure, dielectric
properties, and plasmon excitations in single- and multi-wall carbon nanotubes and carbon onions, for example.
Our preference for plotting the illustrative numerical results in terms of the dimensionless frequency and
propagation vector leaves free an option of choosing the plasma frequency lower or higher, just as the radii
of the cables.

Future dimensions worth adding to the problem remain open in this context. The important issues, which need
to be considered and which could give better insight into the problem, include the role of absorption, the
effects of the spatial dispersion, the coupling to the optical phonons, effect of an applied electric field,
and most importantly the effect of an applied magnetic field in order to study, for example, the edge
magnetoplasmons in the concentric cylindrical cables, to name a few. Choosing the unidentical dielectrics
and/or unidentical metamaterials will only alter the asymptotic limits in the short wavelength limit. Given
a decade of intense research on the metamaterials -- from electromagnetics to acoustics to plasmonics -- it
is now time to devise a serious theory for the inelastic electron and light (or Raman) scattering from the
structures fabricated of metamaterials.

It is by now well-known that proposal of negative refraction burst out as a theoretical concept rather than
an experimental discovery. So, the challenge posed to experiment was to find materials with negative values
of $\epsilon$ and $\mu$. Concurrently, the theory had to defend itself regarding the validity of the concept.
It is fair to say that 2003 marked the beginning of the optimism on the theoretical schemes as well as the
experimental realization of the metamaterials. The real progress made in the subject - both theoretical and
experimental -- during the past decade is remarkable: the opportunities for the response in all frequency
ranges such as rf, microwave, gigahertz, terahertz, and optical region beyond are being exploited rigorously.
The early excitement in electromagnetism has already impregnated the acoustics and plasmonics in the quest of
analogous goals.

Experimental challenge, particularly in the territory of optics and plasmonics, now is to improve the design
of the metamaterials, especially by reducing loss, but also by moving from the laboratory scale to the
industrial scale. While both theory and experiments have made considerable progress in the microwave range,
attempts to explore the THz and visible ranges are obstructed due, in fact, to the inherent losses in the
NIMs. Generally, the loses are orders of magnitude too large for the proposed applications, particularly at
the shorter wavelengths, and the schemes for minimizing such losses with optimized designs do not seem to
have had much luck in the past. A recent proposal to incorporate gain media into NIM designs is reported
to have achieved a significant success in fabricating an extremely low-loss, active optical NIM in order to
probe the optical spectral range [131]. The issue related with the loss is even more solicitous for the
plasmonic phenomena at optical frequencies where the aim is to design devices structured on a subwavelength
scale. By and large, a lossless NIM is still a dream seemingly hard to come true. Giving up is not an option,
however!

\vspace{1.0cm}
%%%%%%%%%%%%%%%%%
{\centerline {\em The greatest glory lies not in never falling, but in rising every time we fall.}}
{\centerline {\hspace{10.7cm} --- Confucius}}
%%%%%%%%%%%%%%%%%

\vspace{0.5cm}
%%%%%%%%%%%%%%%%%%%%%%%%%%%%%%%%%%%%%%%%%%%%%%%%%%%%%%%%%%%%%%%%%%%%%%%%%%%%%%%%%%%%%%%%%%%%%%%%%%%%%%%%%%%%
\begin{acknowledgments}
This project would not have been possible without the support of many people: M.S.K. does not know how to
thank Naomi Halas, Peter Nordlander, and Chizuko Dutta enough who have been so very instrumental behind
his enthusiasm. We also take this opportunity to sincerely acknowledge many very fruitful discussions
with Leonard Dobrzynski. M.S.K. feels thankful to Professor John Pendry and Ms. Cyndie Cumming
[AIP -- Physics Today] for useful communications. He genuinely thanks Kevin Singh for the indispensable
help with the software during the course of this write-up. M.S.K. would also like to thank Professor
T.C. Killian for all the support and encouragement.
\end{acknowledgments}
%%%%%%%%%%%%%%%%%%%%%%%%%%%%%%%%%%%%%%%%%%%%%%%%%%%%%%%%%%%%%%%%%%%%%%%%%%%%%%%%%%%%%%%%%%%%%%%%%%%%%%%%%%%%

\newpage
%%%%%%%%%%%%%%%%%%%%%%%%%%%%%%%%%%%%%%%%%%%%%%%%%%%%%%%%%%%%%%%%%%%%%%%%%%%%%%%%%%%%%%%%%%%%%%%%%%%%%%%%%%%%

%%%%%%%%%%%%%%%%%%%%%%%%%%%%%%%%%%%%%%%%%%%%%%%%%%%%%%%%%%%%%%
%\end{spacing}

\end{document}